\documentclass[12pt]{article}
\usepackage{latexsym,amssymb,amsfonts,graphicx}
\usepackage{amsmath}
\usepackage{amsfonts}
\usepackage{amssymb}
\usepackage{graphicx}
\usepackage{color}
\newcommand{\oo}{\overline}

\newcommand{\ba}{\begin{array}}
\newcommand{\ea}{\end{array}}
\newcommand{\be}{\begin{equation}}
\newcommand{\ee}{\end{equation}}
\newcommand{\bea}{\begin{eqnarray}}
\newcommand{\eea}{\end{eqnarray}}
\newcommand{\beaa}{\begin{eqnarray*}}
\newcommand{\eeaa}{\end{eqnarray*}}

\def\qed{ \hfill \vrule width.25cm height.25cm depth0cm\smallskip}

\newcommand{\basa}{\begin{assumption}}
\newcommand{\easa}{\end{assumption}}

\newcommand{\bas}{\begin{assum}}
\newcommand{\eas}{\end{assum}}

\begin{document}

\begin{center}

{\Large Boolean Delay Equations: \\ 
A Simple Way of Looking at Complex Systems}
\end{center}
\bigskip

Michael Ghil, $^{a,b,c,d}$
\footnote{Corresponding author. Phone: 33-(0)1-4432-2244.
Fax: 33-(0)1-4336-8392.\\ E-mail: ghil@lmd.ens.fr}
Ilya Zaliapin, $^{d,e}$
\footnote{E-mail: zal@unr.edu}
and
Barbara Coluzzi $^b$
\footnote{E-mail: coluzzi@lmd.ens.fr}

\bigskip

\noindent
{\it \small
$^a${D\'{e}partement Terre-Atmosph\`{e}re-Oc\'ean and
Laboratoire de M\'{e}t\'{e}orologie Dynamique
(CNRS and IPSL), Ecole Normale Sup\'{e}rieure, 
24 rue Lhomond, F-75231 Paris Cedex 05, FRANCE.}\\
$^b${Environmental Research and Teaching Institute, 
Ecole Normale Sup\'{e}rieure, F-75231 Paris Cedex 05, FRANCE.}\\
$^c${Department of Atmospheric and Oceanic Sciences,
University of California, Los Angeles, CA 90095-1565, USA.}\\
$^d${Institute of Geophysics and Planetary Physics,
University of California, Los Angeles, CA 90095-1567, USA.}\\
$^e${Department of Mathematics and Statistics,
University of Nevada, Reno, NV 89557, USA.}
}

\bigskip

\begin{abstract}
Boolean Delay Equations (BDEs) are semi-discrete dynamical 
models with Boolean-valued variables that evolve in continuous time. 
Systems of BDEs can be classified into {\it conservative} or {\it dissipative},
in a manner that parallels the classification of ordinary or partial 
differential equations. Solutions to certain conservative BDEs exhibit growth 
of complexity in time. They represent therewith metaphors for biological 
evolution or human history. Dissipative BDEs are structurally stable and 
exhibit multiple equilibria and limit cycles, as well as more complex, fractal 
solution sets, such as Devil's staircases and ``fractal sunbursts.'' All known 
solutions of dissipative BDEs have stationary variance. BDE systems of this 
type, both free and forced, have been used as highly idealized models of 
climate change on interannual, interdecadal and paleoclimatic time scales. 
BDEs are also being used as flexible, highly efficient models of colliding 
cascades of loading and failure in earthquake modeling and prediction, 
as well as in genetics. 
In this paper we review the theory of systems of BDEs and 
illustrate their applications to climatic and solid-earth problems.
The former have used small systems of BDEs, while the latter have used large 
hierarchical networks of BDEs. 
We moreover introduce BDEs with an infinite number of variables distributed 
in space (``partial BDEs'') and 
discuss connections with other types of discrete dynamical systems, including 
cellular 
automata and Boolean networks. This research-and-review paper concludes with a 
set of open questions.
\end{abstract}

\bigskip

\noindent
{\bf Keywords:}
Discrete dynamical systems, Earthquakes, 
El-Ni\~no/Southern-Oscillation,
Increasing complexity, Phase diagram, Prediction.

\smallskip

\noindent
{\bf PACS:}\\
02.30.Ks (Delay and functional equations)\\
05.45.Df (Fractals) \\
91.30.Dk (Seismicity) \\
92.10.am (El Ni\~no Southern Oscillation)\\
92.70.Pq (Earth system modeling).

\section{Introduction}
\label{intro}
BDEs are a modeling framework especially tailored 
for the mathematical formulation of conceptual models of 
systems that exhibit threshold behavior, 
multiple feedbacks and distinct time delays 
\cite{Dee84,Mull84,GhilMull85,GMP87}.
BDEs are intended as a heuristic first step on the way to 
understanding problems too complex to model using systems of 
partial differential equations at the present time. 
One hopes, of course, to be able to eventually write down and solve 
the exact equations that govern the most intricate phenomena. 
Still, in the geosciences as well as in the life 
and other natural sciences, much of the preliminary 
discourse is often conceptual. 

BDEs offer a formal mathematical language that may help bridge 
the gap between qualitative and quantitative reasoning. 
Besides, they are fun to play with and produce beautiful
fractals by simple, purely deterministic rules.
Furthermore, they also provide an unconventional view on the 
concepts of non-linearity and complexity.

In a hierarchical modeling framework, simple 
conceptual models are typically used to 
present hypotheses and capture isolated mechanisms, while more detailed 
models try to simulate the phenomena more realistically and test for the 
presence and effect of the suggested mechanisms 
by direct confrontation with observations \cite{GR00}. 
BDE modeling may be the simplest representation of the relevant 
physical concepts. 
At the same time, new results obtained with a BDE model often 
capture phenomena not yet found by using conventional tools
\cite{SG01,ZKG03a,ZKG03b}. BDEs suggest possible mechanisms 
that may be investigated using more complex models once their 
``blueprint'' is detected in a simple conceptual model.
As the study of complex systems garners increasing attention and 
is applied to diverse areas --- from microbiology to the evolution of 
civilizations, passing through economics and physics --- 
related Boolean and other 
discrete models are being explored more and more 
\cite{Cowan94,Gut91,Wol94,Kau95,GC05}.

The purpose of this research-and-review paper is 
threefold:
(i) summarize and illustrate key properties and applications
of BDEs;
(ii) introduce BDEs with an infinite number of variables; and
(iii) explore more fully connections between BDEs and other
types of {\it discrete dynamical systems} (dDS).
Therefore, we first describe the general form and
main properties of BDEs and place them in the more general
context of dDS, including cellular automata and Boolean
networks (Sect.~\ref{BDE_sect}).
Next, we summarize some applications, to climate dynamics
(Sect.~\ref{ENSO}) and to earthquake physics (Sect.~\ref{CCM});
these applications illustrate both the beauty and usefulness
of BDEs.
In Sect.~\ref{PBDE} we introduce BDEs with an infinite
number of variables, distributed on a spatial lattice
(``partial BDEs'') and point to several ways of
potentially enriching our knowledge of BDEs and extending 
their areas of application.
Further discussion and open questions conclude the paper 
(Sect.~\ref{whatnext}).

\section{Boolean Delay Equations (BDEs)}
\label{BDE_sect} 
\begin{figure}[bp]
\centering\includegraphics[width=.7\textwidth]{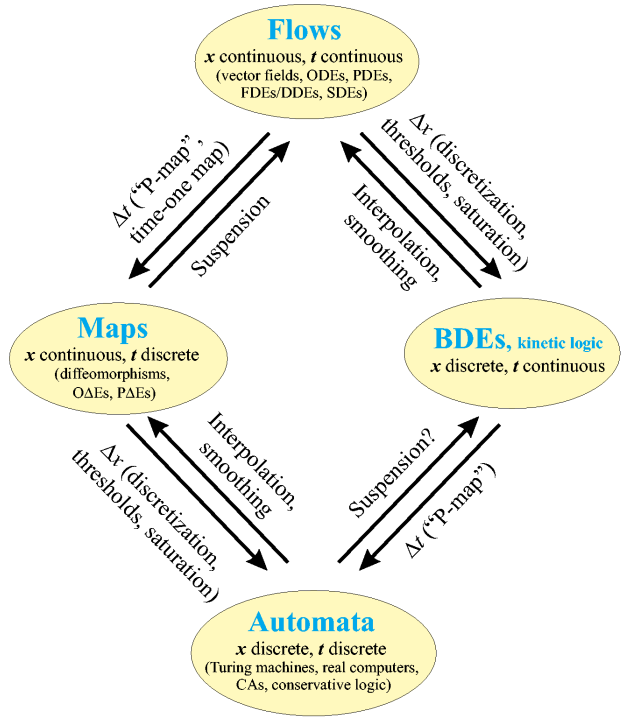}
\caption{The place of BDEs within dynamical system theory.
Note the links: The discretization of $t$ can be 
achieved by the Poincar\'{e} map (P-map) or a 
time-one map, leading from {\bf Flows} to {\bf Maps}. 
The opposite connection is achieved by suspension.
To go from {\bf Maps} to {\bf Automata}
we use the discretization of $x$. 
Interpolation and smoothing can lead in the opposite 
direction.
Similar connections lead from {\bf BDEs} to {\bf Automata}
and to {\bf Flows}, respectively.
Modified after Mullhaupt \cite{Mull84}.
} 
\label{fig_rhomb} 
\end{figure}

BDEs may be classified as {\it semi-discrete dynamical systems}, 
where the variables are discrete --- typically Boolean, 
{\it i.e.} taking the values $0$ (``off'') or  $1$ (``on'') only --- 
while time is allowed to be continuous. 
As such they occupy the previously ``missing corner'' in the 
rhomboid of Fig.~\ref{fig_rhomb}, 
where dynamical systems are classified according 
to whether their time ($t$) and state variables ($x$) are 
continuous or discrete. 

Systems in which both variables and time are 
continuous are called {\it flows} \cite{Arn83,Smale67}
(upper corner in the rhomboid of Fig.~\ref{fig_rhomb}).  
Vector fields, ordinary and partial differential equations (ODEs and PDEs), 
functional and delay-differen\-tial equations (FDEs and DDEs) and stochastic 
differential equations (SDEs) belong to this category. 
Systems with continuous variables and discrete time (middle left corner) 
are known as {\it maps} \cite{Coll80,Hen66} and include diffeomorphisms, 
as well as ordinary and partial difference equations 
(O$\triangle$Es and P$\triangle$Es). 

In automata (lower corner) both the time and the variables
are discrete; 
cellular automata (CAs) and all Turing machines 
(including real-world computers) are part of this group \cite{Gut91,Wol94,vN},
and so is the synchronous version of
Boolean random networks \cite{Kau95,Kau93}.
BDEs and their predecessors, kinetic \cite{Thom79} 
and conservative logic, complete the 
rhomboid in the figure and occupy the remaining middle right corner.

The connections between flows and maps are fairly well understood,
as they both fall in the broader category of {\it differentiable 
dynamical systems} (DDS \cite{Arn83,Smale67,Coll80}).
Poincar\'e maps (``P-maps'' in Fig.~1), which are obtained from
flows by intersection with a plane (or, more generally, with a
codimension-1 hyperplane) are standard tools in the study of 
DDS, since they are simpler to investigate, analytically or
numerically, than the flows from which they were obtained.
Their usefulness arises, to a great extent, from the fact that
--- under suitable regularity assumptions --- the process of
suspension allows one to obtain the original flow from its
P-map; hence the properties of the flow can be deduced from those
of the map, and vice-versa.

In Fig.~1, we have outlined by labeled arrows the processes that 
can lead from the dynamical systems in one corner of the rhomboid 
to the systems in each one of the adjacent corners.
Neither the processes that connect the two dDS corners, automata
and BDEs, nor these that connect either type of dDS with the
adjacent-corner DDS --- maps and flows, respectively ---
are as well understood as the (P-map, suspension) pair of
antiparallel arrows that connects the two DDS corners.
We return to the connection between BDEs and Boolean networks 
in Sect. 2.6 below. The key difference between kinetic logic and BDEs is 
summarized in Appendix~A.

\subsection{General form of a BDE system} 
\label{GF}

Given a system with $n$ continuous real-valued state variables  
${{\bf{v}}=(v_1, v_2, \dots, v_n)}$\\$\in {\mathbb{R}}^n$
for which natural thresholds $q_i\in{\mathbb{R}}$ exist, one 
can associate with each variable $v_i\in{\mathbb{R}}$ 
a Boolean-valued variable, 
$x_i\in{\mathbb{B}}=\{0,1\}$, {\it i.e.}, a variable that is either 
"on" or "off,'' by letting
\begin{equation}
x_i=\left\{ 
\begin{array}{cc} 
0,& v_i \leq  q_i\\ 
1,& v_i > q_i 
\end{array}\right.,
\qquad i=1,\dots,n.
\label{threshold}
\end{equation}  
                                                      
The equations that describe the evolution in 
time of the Boolean vector  
${\bf{x}}=(x_1,x_2,\dots,x_n)\in{\mathbb{B}}^n$ due to the 
time-delayed interactions between the Boolean variables  
$ x_i\in{\mathbb{B}}$ are of the form:
\begin{equation}
\left \{ 
\begin{array}{c} 
x_1(t)= f_1\left[\vphantom{\sum^1}t,x_1(t-\theta_{11}),x_2(t-\theta_{12}),\dots,x_n(t-\theta_{1n})\right],\\
x_2(t)= f_2\left[\vphantom{\sum^1}t,x_1(t-\theta_{21}),x_2(t-\theta_{22}),\dots,x_n(t-\theta_{2n})\right],\\
        \quad \vdots          \\
x_n(t)= f_n\left[\vphantom{\sum^1}t,x_1(t-\theta_{n1}),x_2(t-\theta_{n2}),\dots,x_n(t-\theta_{nn})\right]. 
\end{array} \right.
\label{BDE}
\end{equation}

Here each Boolean variable $x_i$ depends on 
time $t$ and on the state of the other variables $x_j$ in the past. 
The functions $f_i:{\mathbb{B}}^n \rightarrow {\mathbb{B}}$, 
$1\leq i\leq n$, are defined via Boolean equations that involve 
logical operators (see Table 1). 
Each delay value $\theta_{ij}\in{\mathbb{R}}$, 
$ 1\leq i,j\leq n $,  is the length of time it takes for a change 
in variable $x_j$ to affect the variable $x_i$. 
One always can normalize delays $\theta_{ij}$ to be within the 
interval $(0,\,1]$ 
so the largest one has actually unit value;
this normalization will always be assumed from now on.

Following Dee and Ghil \cite{Dee84}, Mullhaupt \cite{Mull84},  
and Ghil and Mullhaupt, 
\cite{GhilMull85} we consider in this section only deterministic,
{\it autonomous} systems with no explicit time dependence.
Periodic forcing is introduced in Sect.~\ref{ENSO}, and
random forcing in Sect.~\ref{CCM}.  
In Sects.~2--4 we consider only the case of $n$ finite 
(``ordinary BDEs''), but in Sect.~5 we allow $n$ to be
infinite, with the variables distributed on a regular lattice
(``partial BDEs''). 

\subsection{ Essential theoretical results on BDEs } 

\label{theory}

We summarize here the most important theoretical results from BDE theory; 
their original and complete form appears in \cite{Dee84,Mull84,GhilMull85}.

We start by choosing a proper topology for the study of BDEs.
Denoting by $\mathbb{B}^n[0,\,1]$ the space of Boolean-valued vector 
functions with a finite number of jumps in the interval $[0,1]$:
\[ {\bf{x}}\mid_{[0,1]} \equiv {\bf{x}}(t\,:\,0\le t\le 1), \]
and noting that, $\forall \tau, {\bf{x}}\mid_{[\tau,\tau+1]}$
still belongs to $\mathbb{B}^n[0,\,1]$ apart from a translation
in time, the system (\ref{BDE}) can be considered
as an endomorphism: 
\begin{equation}
\mathcal{F}_f\,:\,\mathbb{B}^n[0,1]\rightarrow\mathbb{B}^n[0,1].
\end{equation}
We wish to extend this endomorphism into one that acts on
the solutions ${\bf{x}}(t)$ of Eq.~\eqref{BDE}:
\begin{equation}
\mathcal{F}_f\,:\,{\bf{x}}\mid _{[t,\,t+1]} \rightarrow 
{\bf{x}}\mid _{[t+1,\,t+2]}. 
\label{endo}
\end{equation}  
Changing the point of view between (\ref{BDE}) and (\ref{endo})
helps us study the dynamical properties of BDEs.
The space $\mathbb{B}^n[0,\,1]$ equipped with Boolean algebra
\cite{Arnold}
and the topology induced by the $L_1$ metric:
\begin{equation}
d({\bf{x}},{\bf{y}}) \equiv \int_0^1 
\sum_{i=1}^n |x_{i}(t)-y_{i}(t)|\,dt,
\end{equation}
is the {\it phase space} on which $\mathcal{F}$ acts; we denote it by $X$.
In coding theory, this metric is often called the {\it Hamming distance}.

In constructing solutions for a given BDE system,
there is a certain similarity with the theory of real-valued
delay-differential equations (DDEs) 
(see \cite{BG82,Driver77,Hale71,McDonald78}),
as well as with that of ordinary difference equations (O$\Delta$Es)
(\cite{BC63,IK66}).

{\bf Theorem 2.1}
{\rm {\bf (Existence and uniqueness)}}
{\it Let ${\bf{x}}\mid_{[0,1]}\in\mathbb{B}^n[0,1]$
 be the initial data of the dDS (\ref{endo}).
Then the equivalent system (\ref{BDE}) has a unique solution for all 
$t\ge 1$ and for an arbitrary $n^2$-vector of
delays $\Theta= (\theta_{ij})\in (0,\,1]^{n^2}$.}

{\it Sketch of Proof:} The theorem can be proved by induction, constructing 
an algorithm
that advances the solution in time and using a lemma that shows the 
number of jumps (between the values 0 and 
1) to be bounded from above in any finite time interval \cite{Dee84}.
Thus the iterates  $\mathcal{F}^k, \: k=1,\dots,K$, stay within
$\mathbb{B}^n[0,1]$ for all finite $K$ and the unique solution
of (\ref{BDE}) is given simply by piecing together the successive
intervals $[0,1],[1,2],\dots,[K,K+1]$, etc.
\qed

{\bf Theorem 2.2}
{\rm {\bf (Continuity)}}
{\it The endomorphism $\mathcal{F}: X\to X$ is continuous for given delays.
Moreover, the endomorphism 
$\mathcal{F}:X\times [0,\,1]^{n^2}\to X\times [0,\,1]^{n^2}$
is continuous, where the space of delays 
$(0,\,1]^{n^2}$ has the usual Euclidean topology.}
\smallskip

At this point, we need to make the critical distinction between
rational and irrational delays.
All BDE systems that possess only {\it rational} delays can be reduced in 
effect to finite cellular automata. 
Commensurability of the delays creates a partition of the time axis 
into segments over which state variables remain constant and whose 
length is an integer multiple of the delays' least common denominator (lcd). 
As there is only a finite number of possible assignments of two values 
to these segments, repetition must occur, and the only asymptotic 
behavior possible is eventual constancy or periodicity in time. 
Thus, we obtain the following

{\bf Theorem 2.3}
{\rm {\bf (``Pigeon-hole'' lemma)}}
{\it All solutions of (\ref{BDE}) with rational delays 
$\Theta\in\mathbb{Q}^{n^2}$ are eventually periodic.}

{\bf Remark.} By ``eventually'' we mean that a finite-length transient
may occur before periodicity sets in. An interesting feature of BDEs
$vs.$ flows or maps, as we shall see, is precisely that such transients
have finite rather than infinite duration, {\em i.e.}, asymptotic
behavior is reached in finite time.
\smallskip

\begin{figure}[ht]
\centering\includegraphics[width=.7\textwidth]{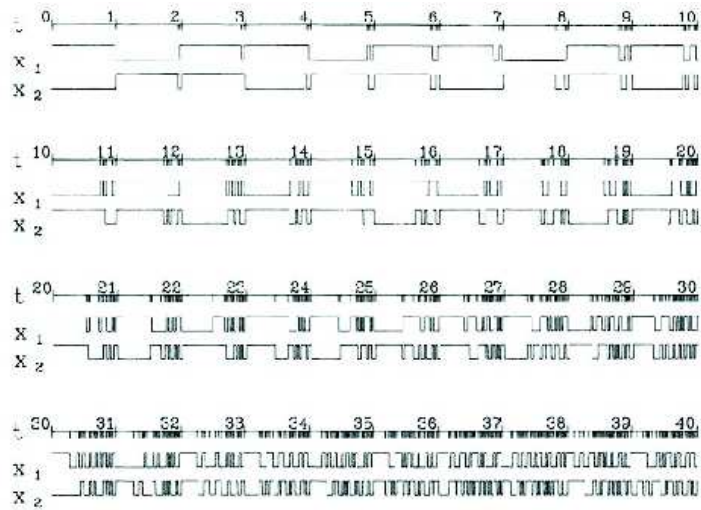}
\caption{Solutions of the system of two conservative BDEs (\ref{DGsyst})
for the delay $\theta=0.977$ and $0\le t<40$ \cite{Dee84}.
The tick marks on the $t$-axis indicate the times at which
jumps in either $x_1$ or $x_2$ take place.
After Dee and Ghil \cite{Dee84}. 
Copyright \copyright Society for Industrial and Applied Mathematics; 
reprinted with permission. 
} 
\label{fig_complex} 
\end{figure}

Dee and Ghil \cite{Dee84}, though, found that for the simple system of 
two BDEs:
\begin{equation}
\left\{ \begin{array}{lcl} 
x_1(t)&=&x_2(t-\theta)\\ 
x_2(t)&=&x_1(t-\theta)\bigtriangledown x_2(t-1),
\end{array}\right.
\label{DGsyst}
\end{equation}  
where $\bigtriangledown$ is the exclusive OR (see Table~1), 
the number of jumps per unit time seemed to keep
increasing with time (see Fig.~\ref{fig_complex})
for a rational value $\theta=0.977$.
Complex, aperiodic behavior only arises in cellular automata
for an infinite number of variables (also called sites).
Thus BDEs seem to pose interesting new problems, irreducible
to cellular automata.
One of these, at least, is the question of which BDEs, if any, 
do posses solutions of increasing complexity.
To answer this question, we need to classify BDEs and to study
separately the effects of rational and irrational delays.

\subsection{Classification of BDEs}

Based on the pigeon-hole lemma, and therefore on the behavior 
for rational delays, Ghil and Mullhaupt \cite{GhilMull85} classified 
BDE systems as follows. 
All systems with solutions that are immediately periodic, for any 
initial data, are \emph{conservative}; 
all other systems are \emph{dissipative} and for some initial data will 
exhibit transient behavior before settling into eventual periodicity or 
quasi-periodicity. 
The DDS analogs are conservative 
({\it e.g.}, Hamiltonian) dynamical systems \cite{Guck83,LL83} versus 
forced-dissipative systems ({\it e.g.}, the well-known Lorenz 
system \cite{Lor63}).
Typical examples of conservative systems occur in
celestial mechanics \cite{Arn78,VRG}, while dissipative
systems are often used in modeling geophysical \cite{GC87,DG05}
and many other natural phenomena.

The simplest nontrivial examples of a conservative and a 
dissipative BDE are
\[x(t) = \oo{x}(t-1)\]
and
\[x(t) = x(t-1)\wedge x(t-\theta), \quad 0<\theta<1,\]
respectively. The Boolean operators we use are listed in Table 1.
It is common to call a Boolean function $f=(f_1,\dots,f_n)$
a {\it connective} and its arguments $x_i$ {\it channels}
\cite{Arnold}; we shall also refer to a channel $x_i$
simply as channel $i$.

{\bf Definition 2.1}
{\it A BDE system is {\emph{conservative}} for an open set
$\Omega\subset(0,\,1]^{n^2}$ of delays if for all rational
delays in $\Omega$ and all initial data there are no 
transients; otherwise the system is {\emph{dissipative}}.}
\label{con}

As is also the case in DDS theory, the conservative
character of a BDE
is tightly connected with its time {\it reversibility}.

{\bf Definition 2.2}
{\it A BDE system is {\emph{reversible}} if its time reversal
also defines a system of BDEs.}
\label{rev}

{\bf Theorem 2.4}
{\rm {\bf (Conservative $\Leftrightarrow$ Reversible) }}
{\it Definitions {\bf 2.1} and {\bf 2.2}  are equivalent.}

Useful algebraic criteria have been established 
\cite{Mull84,GhilMull85} for 
{\it linear} or {\it partially linear} systems of BDEs
to be conservative.
Consider the following system
\begin{equation}
x_i(t)=\sum_{j=1}^n c_{ij}\,x_j(t-\theta_{ij})
\oplus g_i[x_{j'}(t-\theta_{ij'})],
\quad 1\le i\le n;
\label{pl}
\end{equation} 
here $\oplus$ and the summation symbol stand for addition (mod 2) in $X$; 
$c_{ij}\in\mathbb{Z}_2$, where $\mathbb{Z}_2$ is the
{\it field} $\{0,1\}$ associated with this addition, while the
$g_i$ depend only on those $x_{j'}$ for which $c_{ij'}=0$.
Note that
$x\bigtriangledown y = x\oplus y$, while
$x\bigtriangleup y = 1\oplus x\oplus y$.
We use the two types of symbols, $\bigtriangledown$ and 
$\oplus$, interchangeably, depending on the context or point of view.

Adding constants $c_{i0}$ to the above equations
corresponds to adding particular ``inhomogeneous''
solutions to the homogeneous linear system.
All solutions of the full system can be represented
as the sum of solutions to inhomogeneous and homogeneous
systems. We review below only the homogeneous case.

We call a system {\it linear} if and 
only if (iff) all $g_i=0$.
Naturally, the system obtained by putting $g_i=0$ in
(\ref{pl}) is called the {\it linear part} of the BDE system.
Note that this concept of linearity (mod 2) is actually very
nonlinear over the field of reals $\mathbb{R}$, with usual
addition and multiplication: it corresponds, in a sense,
to the thresholding involved in Eq.~(\ref{threshold}).

First we consider the simplest case of systems with distinct 
rational delays in their linear part.
With any such system we associate its {\it characteristic
polynomial}
\begin{equation}
Q(z) = \det A(z),\quad A_{ij}=\delta_{ij}+c_{ij}z^{p_{ij}},
\quad p_{ij}=q\,\theta_{ij},
\label{charp}
\end{equation}
where $q$ is the lcd of all the delays 
$\theta_{ij}$
such that $c_{ij}\ne 0$; the degree of $Q$ is denoted by $\partial Q$.

{\bf Theorem 2.5}
{\rm {\bf (Conservativity for linear systems with distinct 
rational delays)}}
{\it A linear system of BDEs is conservative for an open neighborhood
$\Omega$ of a fixed vector of distinct rational delays 
$\Theta$ iff \[\sum_i q\,\sup_k\,\theta_{ki}=\partial Q.\]}

In the case of rational delays only, we can give a first definition
of {\it partial linearity}, namely that at 
least one $g_i\ne 0$ and $\partial Q\ge 2$.

{\bf Corollary 2.1}
{\rm {\bf (Partially linear systems)}}
{\it The same result holds for a partially linear system of BDEs with
distinct rational delays.}

\subsection{Solutions with increasing complexity}
A natural question is whether (eventually) periodic solutions are generic
in a BDE realm? 
We already noticed (see Fig.~2) that the answer to this question could be 
negative.
Let us introduce the jump counting function
$$J(t)=\#\{ {\mbox{jumps~of~}}{\bf{x}}(t){\mbox{~within~the~interval~}} [t,\,
{t}+1)\},$$
which measures the complexity of a BDE solution ${\bf{x}}(t)$
with a given set of initial data.

{\bf Lemma 2.1}
{\rm {\bf (Increasingly complex solutions for linear BDEs)}}
{\it All solutions (except the trivial one $x(t)\equiv 0$)
of the linear scalar BDE
\begin{equation}
x(t)=x(t-1)\bigtriangledown x(t-\theta_2)\bigtriangledown 
\dots \bigtriangledown x(t-\theta_{\delta})
\label{complex}
\end{equation}
with rationally independent $0<\theta_{\delta}<\dots<\theta_2<\theta_1=1$
and $\delta\ge 2$ are aperiodic and such that the lower bound for the 
corresponding $J(t)$ increases with time.}

\begin{figure}[ht]
\centering\includegraphics[width=.7\textwidth]{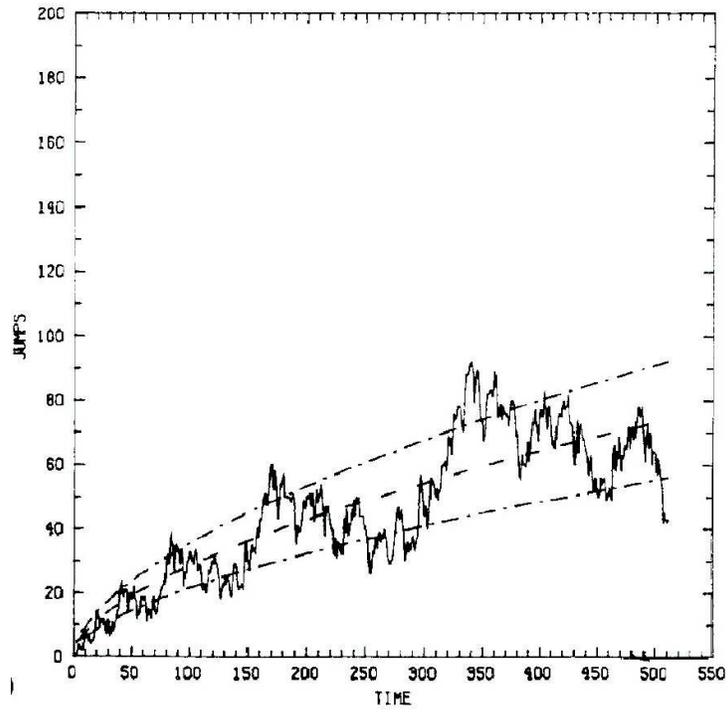}
\caption{Jump function $J(t)$ for the particular solution of
a conservative BDE with the irrational delay 
$\theta=(\sqrt{5}-1)/2$; see Eq.~(\ref{complex}) and further details in the 
text. Reproduced from \cite{GhilMull85}, with kind permission of 
Springer Science and Business Media.} 
\label{fig_complex2} 
\end{figure}

A simple example of this increasing complexity is given in
Fig.~\ref{fig_complex2}, for $\delta=2$, $\theta_2\equiv\theta=(\sqrt{5}-1)/2$,
and a single jump in the initial data.
Note that this delay is equal to the ``golden ratio,''
which is the most irrational number in the sense that
its continued fraction expansion has the slowest possible
convergence \cite{Khinchin}.

{\bf Remark.}
As for ODEs, a ``higher-order'' BDE can easily be written as a set
of ``first-order'' BDEs (\ref{BDE}). 
Therefore the previous lemma also applies to the system (\ref{DGsyst}) 
of two linear BDEs, showing that the complexity of the solution is 
really increasing with time at least for irrational $\theta$.

A more general result holds for {\it partially linear} systems
that include {\it irrational delays}.
For such systems of the form (\ref{pl}), 
we introduce a generalized characteristic polynomial (GCP):
\[Q(\lambda)=\det\,A(\lambda),\quad
A_{ij}(\lambda)=\delta_{ij}+c_{ij}\lambda^{{\theta}_{ij}}.\]
Clearly, this polynomial reduces to 
the characteristic polynomial in (\ref{charp})
if all the delays are rational and $\lambda=z^q$.
The {\it index} $\nu$ of the GCP is defined as the number of its terms. 
We say that a BDE system (\ref{pl}) is {\it partially linear} if
at least one $g_i \neq 0$ and $\nu$ is large enough, $\nu\ge 3$. 

{\bf Theorem 2.6}
{\rm {\bf (Increasingly complex solutions for partially linear BDEs)}}
{\it A partially linear system of BDEs has aperiodic solutions
of increasing complexity, {\it i.e.} with increasing $J(t)$, if its
linear part contains $\delta\ge 2$ rationally independent delays.}
\smallskip

The condition in this theorem is sufficient, but not
necessary.
A simple counterexample is given by the third-order scalar BDE
\begin{equation}
x(t)=\left[x(t-1)\bigtriangledown x(t-\theta)\right]\wedge \oo{x}(t-\tau ),
\end{equation}
with $\theta$, $\tau$, and $\theta/\tau$ irrational, and a single jump
in the initial data at $t_0$: $0<1-\theta<1-\tau<t_0<1$.
The jump function for this solution grows in time like that
of Eq.~(\ref{complex}) for $\delta=2$, although the GCP is identically 1,
so that its index is $\nu=1$.

On the other hand, there exist nonlinear BDE systems with arbitrarily
many incommensurable delays that have only periodic solutions.
For example, all solutions of 
\begin{equation}
\oo{x}(t)=\prod_{k=1}^n x(t-\theta_k)
\label{prod}
\end{equation}
are eventually periodic, with period $\pi=\sum\,\theta_k$ for
$n$ even, and $\pi=2\sum\,\theta_k$ for $n$ odd;
the length $\lambda$ of transients is bounded by $\lambda\le \pi$. 
The multiplication in (\ref{prod}) is in the sense of the field
$\mathbb{Z}_2$, with $xy\equiv x\wedge y$ (see Table 1). 

Dee and Ghil \cite{Dee84} established the upper bound on the jump function,
$J(t)\le K\,t^{l-1}$, where $l$ is, in general, the number of distinct delays 
and the constant $K$ depends only on the vector of delays $\Theta$. 
This bound is essential in proving the existence and uniqueness theorem in 
Sect.~\ref{theory}. 
Moreover, Ghil and Mullhaupt \cite{GhilMull85} obtained
the lower bounds $J(t)=O\left(t^{\log_2(\delta+1)}\right)$
for Eq.~(\ref{complex}) and $J(t) \ge K't^{\log_2\nu}$ 
for partially linear BDEs with $\delta\ge\nu-1$
rationally independent delays in the linear 
part.
These authors also showed the log-periodic character of the jump function
in Fig.~\ref{fig_complex2} (see also Fig.~7 in \cite{GhilMull85}).

Having summarized these results, we are still left with
the question why Fig.~\ref{fig_complex} here, with
$\theta=0.997$ being a rational number, does exhibit increasing
complexity?
The question is answered by the following ``main approximation
theorem''. 

{\bf Theorem 2.7}
{\rm {\bf (Periodic approximation) }}
{\it All solutions to systems of BDEs can be approximated 
arbitrarily well (with respect to the $L_1$-norm of $X$), for a given 
finite time, by the periodic solutions of a nearby system 
that has rational delays only.}
\smallskip

The apparent paradox is thus solved by taking into account
the length of the period obtained for a given conservative BDE
and a given rational delay.
As the lcd $q$ becomes larger and larger, the solution in
Fig.~\ref{fig_complex2} here is well approximated for longer
and longer times (see Fig.~9 of \cite{GhilMull85}); {\it i.e.},
the jump function can grow for a longer time, before 
periodicity forces it to decrease and return to a very small 
number of jumps per unit time.

Since the irrationals are metrically pervasive in  $\mathbb{R}^n$, 
{\it i.e.}, they have measure one, it follows that our chances 
of observing solutions of conservative BDEs with infinite --- or, 
by the approximation theorem, arbitrarily long --- period are 
excellent. 
In fact, the solution shown in Fig.~2 here was discovered pretty 
much by chance, as soon as Dee and Ghil \cite{Dee84} considered a 
conservative system.

Ghil and Mullhaupt \cite{GhilMull85} studied, furthermore, 
the dependence of period length on the connective $f$ and
the delay vector $\Theta$, as well as the degree of 
intermittency of self-similar solutions with growing complexity.
In the latter case, we can consider each solution as a transient
of infinite length.
As we shall see next, such transients preclude structural stability.

\subsection{Dissipative BDEs and structural stability}
The concept of {\it structural stability} for BDEs is patterned 
after that for DDS.
Two systems on a topological space $X$ are said to be {\it 
topologically equivalent} if there exists a homeomorphism 
$h:X\to X$ that maps solution orbits from one system to those 
of the other.
The system is {\it structurally stable} if it is topologically
equivalent to all systems in its neighborhood \cite{Smale67,AP37}. 

In discussing structural stability, we are interested in small
deformations of a BDE leading to small deformations in its solution.
A BDE can be changed by changing either its connective $f$ or
its delay vector $\Theta$.
Changes in $f$ have to be measured in a discrete topology 
and cannot, therefore, be small.
It suffices thus to consider small perturbations of the delays. 

{\bf Theorem 2.8}
{\rm {\bf (Structural stability) }}
{\it A BDE system is structurally stable iff all 
transients and all periods are bounded over 
some neighborhood $U\subset\mathbb{R}^{n^2}$ of 
its delay vector $\Theta$. }
\smallskip

The periodic approximation theorem (Theorem 2.7) implies that,
for BDEs like for DDSs, conservative systems 
are not structurally stable in $X\times [0,1]^{n^2}$. 
Moreover, the conservative ``vector fields,'' here as there, are
in some sense ``rare''; for BDEs they are just the three
connectives $\oo x$, $x\bigtriangledown y$, and $x\bigtriangleup y$,
for which the number of 0's equals the number of 1's 
in the ``truth table.''
Incidentally, the jump set on the delay lattice (see Figs.~1 and 3
of \cite{GhilMull85}), and
hence the growth of $J(t)$, is exactly the same when
replacing $f(x,y)=x\bigtriangledown y$ by
$f(x,y)=x\bigtriangleup y$ \cite{GhilMull85}.

The structural instability and the rarity of conservative BDEs 
justifies studying in greater depth {\it dissipative} BDEs.
Ghil and Mullhaupt \cite{GhilMull85} concentrated on
the scalar $n$th-order BDE
\begin{equation}
x(t)=f[x(t-\theta_1),\dots,x(t-\theta_n)].
\label{scalar}
\end{equation}
The connective $f$ is most conveniently expressed
in its {\it normal forms} from switching and automata theory, with
$xy=x\wedge y$ and $x+y = x\vee y$.
With this notation, the {\it disjunctive} and {\it conjunctive} normal
forms represent $f$ as a sum of products and a product
of sums, respectively.
This formalism helps prove that certain BDEs of the 
form (\ref{scalar}) lead to {\it asymptotic simplification},
{\it i.e.}, after a finite transient, the solution of
the full BDE satisfies a simpler BDE.
An illustrative example is 
\begin{equation}
x(t)=x(t-\theta_1)\,\oo x(t-\theta_2),
\label{illustr}
\end{equation} 
where either $\theta_1$ or $\theta_2$ can be the larger
of the two.
Asymptotically, the solutions of Eq.~(\ref{illustr})
are given by those of a simpler equation 
\[x(t)=x(t-\theta_1).\]

Comparison with the asymptotic behavior of forced-dissipative
systems in the DDS framework shows two advantages of BDEs.
First, the asymptotic behavior sets in after finite
(rather than infinite) time.
Second, the behavior on the ``inertial manifold''
or ``global attractor'' here can be described explicitly
by a simpler BDE, while this is rarely the case
for a system of ODEs, FDEs, or PDEs.

Finally, one can study asymptotic stability of solutions
in the $L_1$-metric of $X$.
We conclude this theoretical section by recalling that, for
$0<\theta <1$ irrational, the solutions of
\[x(t)=x(t-\theta)\,x(t-1)\]
are eventually equal to $x(t)\equiv 0$, except for $x(t)\equiv 1$,
which is unstable.
Likewise, for
\[x(t)=x(t-\theta)+x(t-1),\]
$x(t)\equiv 1$  
is asymptotically stable, while
$x(t)\equiv 0$ is not. More generally, one has the following

{\bf Theorem 2.9}
{\it Given rationally unrelated delays $\Theta=\{\theta_k\}$,
the BDE
\[x(t)=\prod_{k=1}^n x(t-\theta_k)\]
has $x(t)\equiv 0$ as an asymptotically stable solution,
while for the BDE
\[x(t)=\sum_{k=1}^n x(t-\theta_k),\]
$x(t)\equiv 1$ is asymptotically stable.}

To complete the taxonomy of solutions, we also note
the presence of {\it quasi-periodic} solutions;
see discussion of Eq.~(6.18) in Ghil and Mullhaupt 
\cite{GhilMull85}.

{\bf Asymptotic behavior.} 
In summary, the following types of asymptotic behavior were 
observed and analyzed in BDE systems: 
\emph{(a) fixed point} --- the solution reaches one of a 
finite number of possible states and remains there; 
\emph{(b) limit cycle} --- the solution becomes periodic; 
\emph{(c) quasi-periodicity} --- the solution is a sum of
several incommensurable ``modes''; and 
\emph{(d) growing complexity } --- the solution's number of 
jumps per unit time increases with time. 
This number grows like a positive, but fractional power
of time $t$ \cite{Dee84,Mull84}, with superimposed log-periodic
oscillations \cite{GhilMull85}.

\subsection{BDEs, cellular automata (CAs) and Boolean networks}

We complete here
the discussion of Fig.~1 about the place of
BDEs in the broader context of dynamical systems in general.
Specifically, we concentrate on the relationships between BDEs
and other dDS, to wit cellular automata and Boolean networks.

The formulation of BDEs was originally inspired by advances
in theoretical biology, following Jacob and Monod's discovery 
\cite{JacMo61} of on-off interactions between genes, which had 
prompted the formulation of ``kinetic logic'' \cite{Thom79,Thom73,Thom78} 
and Boolean regulatory networks \cite{Kau95,Kau93,Kau69}.
In the following, we briefly review the latter and discuss their relations 
with systems of BDEs, whereas kinetic logic is touched upon in 
Appendix~A.

In order 
to understand the links between BDEs and Boolean regulatory
networks it is important to start by recalling some well known definitions
and results about cellular automata (CAs), which were introduced
by von Neumann already in the late 1940s \cite{vN}. 
Doing so here will also facilitate the discussion of our
preliminary results on ``partial BDEs'' in Sect.~\ref{PBDE}.

One defines
a CA as a set of $N$ Boolean variables $\{ x_i: \: i=1,\dots,N \}$ on the 
sites of a regular lattice in $D$ dimensions. 
The variables are usually updated 
synchronously according to the same deterministic rule 
$x_i(t)=f[x_i(t-1), \dots, x_N(t-1)]$; 
that is the value of each variable $x_i$ at epoch $t$ is determined by the 
values of this and possibly some other variables $\{x_j\}$ 
at the previous epoch $t-1$. 
In the simplest case of $D=1$ ({\em i.e.}, of a 1-D lattice) and 
first-neighbor interactions only, there 
are $2^8$ possible rules $f: \mathbb{B}^3\rightarrow \mathbb{B}$, which give 
256 different elementary CAs 
(ECAs) studied in detail by Wolfram \cite{Wol94,Wol83}. For a given $f$, 
they evolve according to:
\begin{equation}
x_i(t)=f[x_{i-1}(t-1),x_i(t-1),x_{i+1}(t-1)], \hspace{.3in} 1 \le i \le N,
\label{ECA}
\end{equation}
For a finite size $N$, Eq.~(\ref{ECA}) is
a particular case of a BDE system (\ref{BDE}) with connective 
$f_i=f$ for all $i$ and a single delay 
$\theta_{ij}=1$ for all $i$ and $j$ (see also Sect.~\ref{PBDE}). 
One generally speaks of asynchronous CAs when variables at different sites are 
updated at different $discrete$ times according
to some deterministic scheme. Such asynchronous CAs still belong to a 
restricted class of BDEs with integer delays $\theta_{ij} \in \mathbb{N}$.

When both the space and the time are discrete, a finite-size 
CA will ultimately display either a fixed-point or periodic behavior. 
An important advantage of the great simplicity of ECAs is that it allows for 
systematic studies and helps
understand their behavior in the limit of $N \rightarrow \infty$. 
It can be shown that different updating rules can lead to very 
different long-time dynamics. Wolfram \cite{Wol83,Wol84} divided ECAs into 
four universality classes, according to 
the typical behavior observed for 
random initial states and large sizes $N$: 
For rules in the first class the system evolves towards a fixed point.
For rules in the second class the dynamics can attain
either a fixed point or a limit cycle, but
in this case the period is usually small and it remains small for
increasing $N$-values. 
For rules in the third class, though,
the period of the limit cycle usually increases with the size $N$ and it
can diverge in the limit $N\rightarrow\infty$, leading to ``chaotic''
behavior. 
Finally, CAs in the fourth class are capable of 
universal computation and are thus equivalent to a Turing machine.

A first generalization of CAs are Boolean networks, in which the 
Boolean variables $\{ x_i: \: i=1,2,\dots,N \}$ are attached to the nodes 
(also called vertices) 
of a (possibly directed) graph and they evolve synchronously according to 
deterministic Boolean rules, which may vary from node to node.  
A further generalization is obtained by considering randomness, in the 
connections and/or in the choice of updating rules. 
In particular, the $NK$ model introduced by Kauffman \cite{Kau93,Kau69}, 
is among the most extensively analyzed random Boolean networks (RBNs). 
This model considers a system of $N$ Boolean variables such that each value 
$x_i$ 
depends on $K$ randomly chosen other variables $x_j$ through a Boolean 
function 
drawn randomly and independently from ${2^2}^K$ possible variants.
The connections among the 
variables and the updating functions are fixed during 
a given system's evolution, and one looks
for average properties at long times. 
Since the variables are updated synchronously, at the same $discrete$ 
$t$-values, 
the evolution will ultimately reach a fixed point or a limit cycle for any 
given configuration of links and rules. 

Kauffman \cite{Kau93,Kau69} proposed such $NK$ RBNs as models of a 
regulatory genetic network, with different nodes corresponding to different 
genes. 
The activity of a gene $x_i$ is regulated by the activity of the other $K$ 
genes 
to which $x_i$ is connected. 
The different attractors, whether fixed point or limit cycle, are 
related to different gene expression patterns. In this interpretation, 
a limit cycle, {\em i.e.} a recurrent pattern, corresponds to a 
cell type and the period is that of the cell cycle.

The $NK$ model was initially studied for a uniform distribution of the 
updating rules. 
In this situation, for small $K$ values, one 
finds on average a small number of fixed points and limit cycles. 
The lengths of the possible attractors remain finite in the limit of 
$N \rightarrow \infty$, 
and therefore the network dynamics appears ``ordered.'' 
For large $K$ values, though, the model displays ``chaotic'' behavior; 
in this case, the average number of attractors as well as their average 
length diverge with $N$ and the difference between two almost identical 
initial states can increase exponentially with time. 
Furthermore, one observes a transition in parameter space between typical 
dynamics that is characterized by a large connected cluster of frozen 
variables and the opposite one with small separated clusters of frozen 
variables. The critical values of the parameters corresponding to this passage 
from an ``ordered'' to a ``chaotic'' regime can be evaluated by looking at the 
evolution of the Hamming distance between two trajectories that start 
from slightly different configurations 
\cite{We91,DePo86,DeWe86}. In particular, for a uniform distribution of 
the Boolean updating functions, the $NK$ model is ``critical'' when $K=2$.

Kauffman \cite{Kau93} suggested that natural organisms could lie on or near 
the borderline between these two different dynamical regimes, {\it i.e.} 
``at the edge of chaos,'' where the system is still sufficiently robust
against small perturbations but at the same time close enough
to the chaotic regime to feel the effect of selection.
Accordingly, a lot of attention has been devoted to the study of such 
critical networks, which can also be obtained for $K>2$ with appropriate,
and possibly more realistic, choices of the updating rule distribution. 
Similar edge-of-chaos suggestions have been made in other applications
of dynamical systems theory, including DDS and celestial mechanics
\cite{VGK}.

For large $N$-values, even the problem of determining 
the fixed points of a generic regular RBN, with $K_i=K$ for all $i$,
is highly nontrivial. In the context of the modeling of genetic interactions, 
the solution to this problem is thought to represent different accessible 
states of the cell, possibly triggered by external inputs 
\cite{CoLePaWeZe06a}. This problem has been recently reformulated 
\cite{CoLePaWeZe06a,CoLePaWeZe06b} in terms 
of the zero-energy configurations of an appropriate Hamiltonian.
In this formulation, statistical mechanics tools from spin-glass physics
can be brought to bear on the problem; these tools have also been 
successfully extended to general optimization issues \cite{MePaZe02}. 

Irregular RBNs, in which the number of inputs $K_i$ is also a node-dependent 
random variable, are obviously harder to analyze. 
There is increasing evidence \cite{AlBa02} that many
networks arising in very different natural contexts are ``scale free,'' 
{\it i.e.} their node-dependent connectivity $K_i$ is distributed according 
to a power law $\mathsf{P}(K_i) \propto {K_i}^{-\gamma}$. 
This seems to be true as well for the distribution 
of the input connections of some genetic networks. 
In the irregular case, too, one still observes ``critical'' dynamical 
behavior, given a suitable distribution of the updating Boolean 
functions. Kauffmann and colleagues \cite{KaPeSaTr04} have recently studied
the stability properties of regular and irregular RBNs and their dependence on 
the distribution of connectivity $K_i$ and/or Boolean functions.
{\it Inverse problems}, in which one tries to determine the Boolean rules 
leading 
to a particular type of behavior, have been considered in \cite{MeTe05}.

The dependence of the average number $\bar{m}$ of attractors and of their 
period length 
on the size $N$ in critical RBNs is still a matter of debate. 
This issue is particularly relevant for genetic-network modeling, since the 
behavior of $\bar{m}(N)$ is expected to be related \cite{Kau93,We91} to
 the number of cell types which are present in an organism characterized by 
a given number $N$ of genes. 
Recent results \cite{SaTr03,DrMiGr05,MiDr06} show that
$\bar m(N)$ increases faster than any power of $N$ in regular synchronous 
RBNs, in which all variables are updated in parallel at the same discrete epochs. 
These findings suggest that using different updating schemes could lead to more 
realistic behavior, with a slower increase of $\bar m(N)$.
Moreover, assuming that all the variables act synchronously, 
{\it i.e.} that they ``move in lock-step,'' may be too drastic a simplification for 
correctly modeling a number of natural systems and, in particular, 
interacting 
genes \cite{Ge04}. In order to overcome this simplification, asynchronous
Boolean networks with different updating procedures have been studied
recently \cite{Ge02}. A recent review  \cite{Ge04} underlines the links 
of such generalized RBNs with 
asynchronous CAs and with ``kinetic logic.'' 

From the point of view of connections with BDEs,  
the updating scheme introduced by Klemm and Bornholdt \cite{KlBo05} is of 
particular interest; they consider a ``critical'' regular RBN with $K=2$ and  
weakly fluctuating delays in the response of each node. 
The number of stable attractors in this system increases more slowly with 
system size than for synchronous updating. 
It seems therefore that RBNs in continuous time may be more realistic 
and may exhibit new and possibly unexpected types of behavior. 

\"Oktem {\it et al.} \cite{Oktem03} have recently applied a BDE 
approach to a Boolean network of genetic interactions with given architecture. 
In this case continuous time delays are introduced according to the BDE 
formalism of \cite{Dee84,Mull84,GhilMull85,GMP87}. 
As a result, more complicated types of behavior than in  synchronously 
updated Boolean networks have been observed, and the dynamics of the system 
seems to be characterized by aperiodic attractors. 

In both \cite{KlBo05} 
and \cite{Oktem03}, the authors introduced a minimal time interval below 
which changes in a given variable are not permitted. 
Such a cutoff, or ``refractory period'' \cite{Oktem03}, may have a 
physiological basis in genetic applications, but it rules out the presence of 
solutions with increasing complexity. 
Therefore, for a finite number of variables, this restriction must result in 
an ultimately periodic behavior; the asymptotic period, though, could be 
much larger than the one obtainable with usual Boolean networks, especially 
when considering conservative connectives and irrational delays.
From this point of view, the implementation of continuous time delays 
in \cite{KlBo05,Oktem03} is different than in BDEs 
\cite{Dee84}-\cite{GMP87} 
and is similar to the one adopted in 
``kinetic logic'' \cite{Thom79,Thom73,Thom78}, whose precise connections with
our formalism are discussed in Appendix~A. 

In the applications of BDEs that we review in the next 
sections, one finds different mechanisms that lead to aperiodic 
solutions of bounded complexity, without the need of a cutoff; one could 
thus explore the possibility of similar behavior in genetic-interaction
models as well. 
Summarizing, one can say that kinetic logic and the recently 
proposed genetic network models \cite{KlBo05,Oktem03}, as well as others 
recent generalizations of RBNs with deterministic updating \cite{Ge04}, 
can be viewed either as asynchronous CAs or as particular cases of BDEs
with large $N$.

In Sect.~5, we initiate the systematic study of BDE systems in the limit of
an infinite number of variables, assumed for the moment to lie on a
regular lattice and to interact according to a given, unique, deterministic 
rule. 
This study should allow us to better understand the connections of BDEs 
with (infinite) CAs, on the one hand, and with PDEs on the other.
Such a study should also help clarify further the 
behavior of, possibly random, Boolean networks in continuous time.

We now turn to an illustration of BDE modeling in action, first with a 
climatic example and then with one from lithospheric dynamics.
Both of these applications introduce new and interesting properties of
and extensions to BDEs. The climatic BDE model in Sect.~3, while keeping a 
small number of variables, introduces variables with more than two levels, as
well as periodic forcing. Its solutions show that a simple BDE model can
mimic rather well the solution set of a much more detailed model,
based on nonlinear PDEs, as well as produce new and previously 
unsuspected results, such as a Devil's Staircase and a ``bizarre'' attractor
in phase-parameter space.

The seismological BDE model in Sect.~4 introduces a much larger
number of variables, organized in a directed 
graph, as well as random forcing and state-dependent delays.
This BDE model also reproduces a regime diagram of seismic sequences
resembling observational data, as well as the results of much more
detailed models \cite{GKZ+00a,GKZ+00b} based on a large system of 
differential equations; furthermore it allows the exploration of 
seismic prediction methods.

\section{A BDE Model for the El Ni\~no/Southern Oscillation}
\label{ENSO}
BDEs were first applied to paleoclimatic problems.
Ghil {\it et al.} \cite{GMP87} used the exploratory power of BDEs 
to study the coupling of radiation balance of the Earth-atmosphere
system, mass balance of continental ice sheets, and overturning 
of the oceans' thermohaline circulation during glaciation cycles.
On shorter time scales, Darby and Mysak \cite{DM93} and 
Wohlleben and Weaver \cite{WW95} studied the coupling of the
sea ice with the atmosphere above and the ocean below in
an interdecadal Arctic and North Atlantic climate cycle,
respectively.
Here we describe an application to tropical climate, on
even shorter, seasonal-to-interannual time scales.

The El-Ni\~no/Southern-Oscillation (ENSO) phenomenon is the most 
prominent signal of seasonal-to-interannual climate variability. 
It was known for centuries to fishermen along the west 
coast of South America, who witnessed a seemingly sporadic and 
abrupt warming of the cold, nutrient-rich waters that 
support the food chains in those regions; these warmings
caused havoc to their fish harvests \cite{Diaz92,Phil90}. 
The common occurrence of such warming shortly after Christmas inspired them to 
name it El Ni\~no, after the ``Christ child.'' 
Starting in the 1970s, El Ni\~no's climatic effects were found to 
be far broader than just its manifestations off the shores of Peru
\cite{Diaz92,Glantz+91,MK05}.
This realization led to a global awareness of ENSO's significance, 
and an impetus to attempt and improve predictions 
of exceptionally strong El Ni\~no events \cite{Latif94,GJ98}.

\subsection{Conceptual  ingredients}
The following conceptual elements are incorporated into the logical equations 
of our BDE model for ENSO variability.

{\bf \textit{(i) The Bjerknes hypothesis:}} 
Bjerknes \cite{Bjer69}, who laid the foundation 
of modern ENSO research, suggested a {\em {positive feedback}} as 
a mechanism for the growth of an internal instability 
that could produce large positive anomalies of sea surface
temperatures (SSTs) in the eastern Tropical Pacific.
We use here the climatological meaning of the term {\em anomaly},
{\em i.e.}, the difference between an instantaneous (or short-term average)
value and the {\em normal} (or long-term mean).
Using observations from the International Geophysical 
Year (1957-58), he realized that this mechanism must
involve {\em{air-sea interaction}} in the tropics. 
The ``chain reaction'' starts with an initial warming of 
SSTs in the ``cold tongue'' that occupies 
the eastern part of the equatorial Pacific. 
This warming causes a weakening of the thermally direct Walker-cell 
circulation; this circulation involves air rising over the
warmer SSTs near Indonesia and sinking over the
colder SSTs near Peru.
As the trade winds blowing from the east weaken and give way to 
westerly wind anomalies, the ensuing local changes in the ocean 
circulation encourage further SST increase. 
Thus the feedback loop is closed and further amplification of the 
instability is triggered.

{\bf \textit{(ii) Delayed oceanic wave adjustments:}} 
Compensating for Bjerknes's positive feedback is a 
{\em{negative feedback}} in the system that allows 
a return to colder conditions in the basin's eastern part. 
During the peak of the cold-tongue warming, called the 
{\em warm} or {\em El~Ni\~no} phase of ENSO, westerly wind anomalies 
prevail in the central part of the basin. 
As part of the ocean's adjustment to this atmospheric forcing, 
a Kelvin wave is set up in the tropical wave guide and carries 
a warming signal eastward. 
This signal deepens the eastern-basin thermocline,
which separates the warmer, well-mixed surface waters from the colder waters
below, and thus contributes 
to the positive feedback described above. 
Concurrently, slower Rossby waves propagate westward, 
and are reflected at the basin's western boundary, giving rise 
therewith to an eastward-propagating Kelvin wave that has a cooling, 
thermocline-shoaling effect. 
Over time, the arrival of this signal erodes the warm event, 
ultimately causing a switch to a {\em cold}, {\em La~Ni\~na} phase.

{\bf \textit{(iii) Seasonal forcing:} } 
A growing body of work \cite{GR00,Cha94,Cha95,JNG94,JNG96,Tzi+94,Tzi+95} 
points to resonances between the Pacific basin's intrinsic air-sea 
oscillator and the annual cycle as a possible cause for the tendency 
of warm events to peak in boreal winter, as well as for ENSO's 
intriguing mix of temporal regularities and irregularities. 
The mechanisms by which this interaction takes place are numerous 
and intricate and their relative importance is not yet fully 
understood \cite{Tzi+95,Batt88,HD05}. 
We assume therefore in the present BDE model that the 
climatological annual cycle provides for a seasonally varying 
potential of event amplification.

\subsection{Model variables and equations}
The model \cite{SG01} operates with five Boolean variables.
The discretization of continuous-valued SSTs and surface winds into
four discrete levels is justified by the pronounced multimodality
of associated signals (see Fig.~1b of \cite{SG01}). 

The state of the {\it ocean} is depicted by SST 
anomalies, expressed via a combination of two Boolean variables, 
$T_1$ and $T_2$. 
The relevant anomalous {\it atmospheric} conditions in the Equatorial 
Pacific basin are described by the variables $U_1$
and $U_2$.
The latter express the state of the trade winds.
For both the atmosphere and the ocean, the first
variable, $T_1$ or $U_1$, describes the sign of the anomaly,
positive or negative, while the second one, $T_2$ or $U_2$,
describes its amplitude, strong or weak.
Thus, each one of the pairs $(T_1,\,T_2)$
and $(U_1,\,U_2)$ defines a four-level discrete
variable that represents highly positive,
slightly positive, slightly negative,
and highly negative deviations from the climatological
mean.
The {\it seasonal cycle's} external forcing is represented by a 
two-level Boolean variable $S$. 

The atmospheric variables $U_i$ are "slaved" to 
the ocean \cite{JNG96,NeelatJin94}:
\begin {equation}
 U_i(t)=T_i(t-\beta),~i=1,2.
\end{equation}    
The evolution of the sign $T_1$ of the SST
anomalies is modeled according to the following two sets of delayed 
interactions:

(i) Extremely anomalous wind stress conditions are assumed to be necessary 
to generate a significant Rossby-wave signal $R(t)$, which takes on the value 
$1$ when wind conditions are extreme at the time and $0$ otherwise. 
By definition strong wind anomalies (either easterly or westerly) prevail 
when $ U_1=U_2 $ and thus $ R(t)= U_1(t) \bigtriangleup U_2(t) $; here 
$\bigtriangleup$ is the binary Boolean operator that takes on the value $1$ if 
and only if both operands have the same value (see Sect.~\ref{BDE_sect} and 
Table~1). A wave signal $ R(t)=1$ that is elicited at time $t$ is assumed to 
re-enter the model system after a delay $\tau $, associated with the wave's 
travel time across the basin. 
Upon arrival of the signal in the eastern equatorial Pacific at time $t+\tau$, 
the wave signal affects the thermocline-depth anomaly there and thus reverses 
the sign of SST anomalies represented by $T_1$.

(ii) In the second set of circumstances, when $ R(t)=0$, and thus no 
significant wave signal is present, we assume that $T_1(t+\tau)$ responds 
directly to local atmospheric conditions, after a delay $\beta$, 
 according to Bjerknes hypothesis; the delays 
associated with local coupled processes are taken all 
equal. 

The two mechanisms (i) and (ii) are combined to yield:
\begin{equation}
 T_1(t)=\left \{ \left [ R \wedge \oo{U_1} \right ] (t-\tau )\right \} 
\lor \left \{ \oo{R}(t-\tau) \wedge U_2(t-\beta)\right \} ;
\end{equation}    
here the symbols $\lor $ and $\wedge$ represent the binary logical 
operators OR and AND, respectively (see Table 1).

The seasonal-cycle forcing $S$ is given by
$S(t)=S(t-1)$; the time $t$ is thus measured in units of 1 year.
The forcing $S$ affects the SST anomalies' amplitude
$T_2$ through an enhancement of events when favorable seasonal 
conditions prevail:
\begin{equation}
T_2(t)=\left \{ \left [ S \triangle T_1 \right ] (t-\beta) \right \} \lor 
\left \{ \left [ \oo{( S \triangle T_1)}  \land T_2 \right ] (t- \beta ) 
\right \}. 
\end{equation} 

The model's principal parameters are the two delays
$\beta$ and $\tau$ associated with
local adjustment processes and with basin-wide processes,
respectively.
The changes in wind conditions are assumed to lag the SST 
variables by a short delay $\beta$, of the order of days to weeks. 
For the length of the delay $\tau$ we adopt Jin's \cite{J96}
view of the delayed-oscillator mechanism and let it represent the 
time that elapses while combined processes of oceanic adjustment occur:
it may vary from about one month in the fast-wave 
limit \cite{JinNeel93,Jn93b,JN93c} to about two years. 

\subsection{Model solutions}
Studying the ENSO phenomenon, we are primarily interested 
in the dynamics of the SST states, represented by the 
two-variable Boolean vector $(T_1,\,T_2)$.
To be more specific, we deal with a four-level scalar variable
\begin{equation}
ENSO=\left\{
\begin{array}{rcl}
-2,&{\rm~extreme~La~Ni\tilde{n}a},&T_1=0,\,T_2=0,\\
-1,&{\rm~mild~La~Ni\tilde{n}a},&T_1=0,\,T_2=1,\\
1,&{\rm~mild~El~Ni\tilde{n}o},&T_1=1,\,T_2=0,\\
2,&{\rm~extreme~El~Ni\tilde{n}o},&T_1=1,\,T_2=1.\\
\end{array}
\right.
\end{equation}

\begin{figure}[ht]
\centering\includegraphics[width=.7\textwidth]{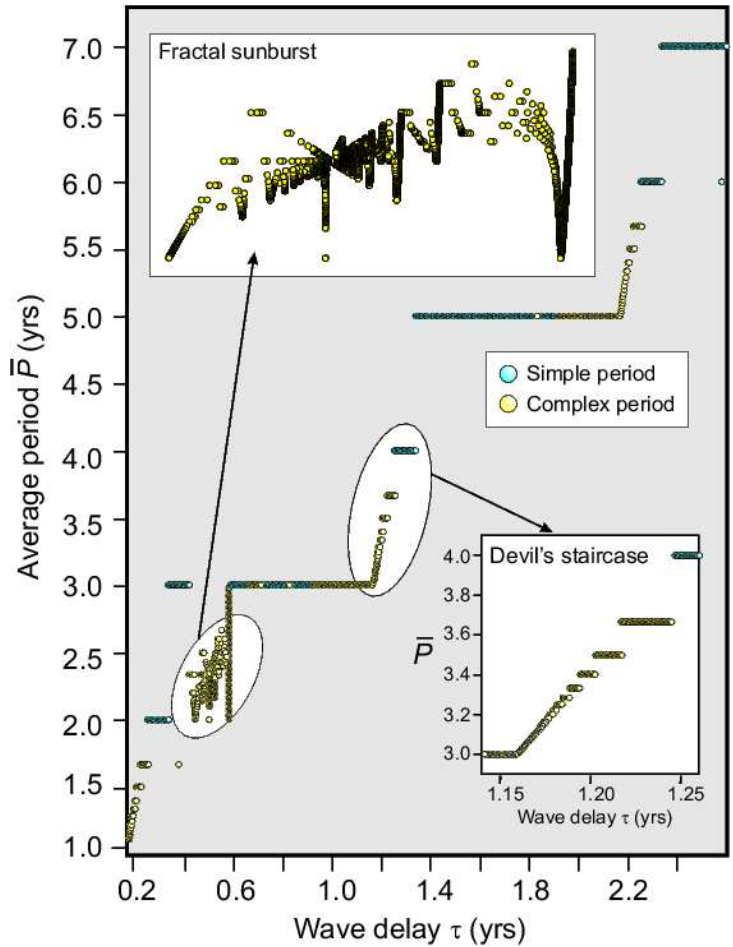}
\caption{Devil's staircase and fractal sunburst for a BDE model of 
the El-Ni\~no/Southern-Oscillation (ENSO) phenomenon. 
Plotted in the bifurcation diagram 
is the average cycle length  $\bar{P}$ vs. 
the wave delay $\tau$  for a fixed local delay $\beta=0.17$. 
Blue dots indicate purely periodic solutions; 
orange dots are for complex periodic solutions; 
small black dots denote aperiodic solutions. 
The two insets show a blow-up of the overall, 
approximate behavior  
between periodicities of two and three years 
(``fractal sunburst'') and of three and four years
(``Devil's staircase''). 
Modified after Saunders and Ghil \cite{SG01}.} 
\label{fig_devil2d} 
\end{figure}

\begin{figure}[ht]
\centering\includegraphics[width=.7\textwidth]{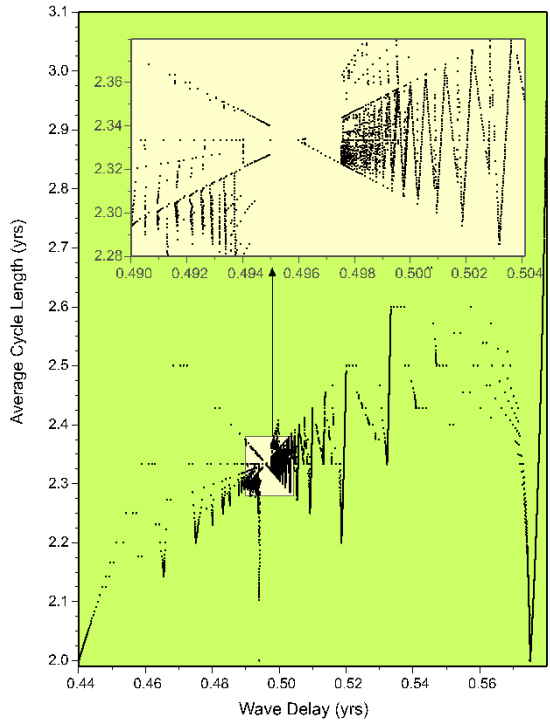}
\caption{Fractal sunburst: a BDE solution
pattern in phase-parameter space, for a dissipative BDE system
with periodic forcing. The plot is
a blow-up of the transition zone from 
average periodicity two to three years in Fig.~\ref{fig_devil2d}; 
here $\tau=0.44$--$0.58, \beta=0.17$. 
The inset is a zoom on $0.490\leq\tau\leq0.504$. 
A complex mini-staircase structure reveals 
self-similar features, with a focal point 
at $\tau\approx 0.5$. 
Modified after Saunders and Ghil \cite{SG01}.} 
\label{fig_burst} 
\end{figure}

\begin{figure}[ht]
\centering\includegraphics[width=.7\textwidth]{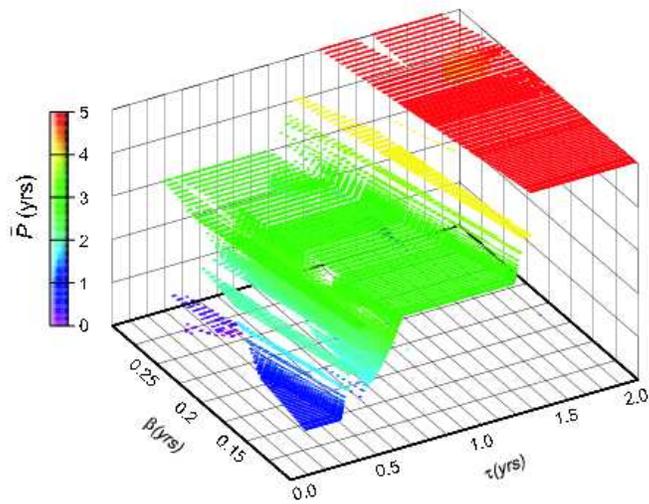}
\caption{The Devil's bleachers in our BDE model of ENSO. The 
three-dimensional regime diagram shows the 
average cycle length $\bar{P}$, portrayed in 
both height and color, vs. the two delays 
$\beta$  and $\tau$. 
Oscillations are produced even for very small 
values of $\beta$, as long as $\beta\leq\tau$. 
Variations in  $\tau$ determine the oscillation's 
period, while changing $\beta$ establishes the 
bottom step of the staircase, shifts the 
location of the steps, and determines their width.
After Saunders and Ghil \cite{SG01}.} 
\label{fig_devil3d} 
\end{figure}

In all our simulations, this variable
takes on the values $\{-2,\,-1,\,1,\,2\}$, precisely in 
this order, thus simulating real ENSO cycles.
The cycles follow the same sequence of states, 
although the residence time within each state changes 
as $\tau$ changes at fixed $\beta$.
The period $P$ of a simple oscillatory solution is 
defined as the time between the onset of two 
consecutive extreme warm events, $ENSO=2$. 
We use the cycle period definition to classify 
different model solutions (see Figs.~\ref{fig_devil2d}--\ref{fig_devil3d}).

{\bf \emph{(i) Periodic solutions with a single cycle (simple period).} }  
Each succession of events, or \emph{internal cycle}, is 
completely phase-locked here to the seasonal cycle, {\it i.e.}, the 
warm events always peak at the same time of year. 
For each fixed $\beta$, as  $\tau$ is increased, 
intervals where the solution has a simple period equal 
to $2, 3, 4, 5, 6$ and $7$ years arise consecutively. 

{\bf \emph{(ii) Periodic solutions with several cycles (complex period).}} 
We describe such sequences, in which several distinct cycles
make up the full period,
by the parameter $\bar{P}=P/n$; here  
$P$ is the length of the sequence and $n$  
is the number of cycles in the sequence. 
Notably, as we transition from a period of three years to a period of 
four years (see second inset of Fig.~\ref{fig_devil2d}),  
$\bar{P}$ becomes a nondecreasing step function of $\tau$ that takes 
only rational values, arranged on a Devil's Staircase. 

\subsection{The quasi-periodic (QP) route to chaos in the BDE model}
The frequency-locking behavior observed for our BDE solutions above 
is a signature of the universal QP route to chaos. 
Its mathematical prototype is the Arnol'd circle map  \cite{Arn83}, 
given by the equation
\begin{equation}
\label{arnold}
\theta_{n+1}=\theta_{n}+\Omega+2{\pi}K\sin(2\pi\theta_n)
~({\rm mod~} 1).
\end{equation}                                        
Equation (\ref{arnold}) describes the motion of a point, denoted 
by the angle $\theta$ of its location on a unit circle, which undergoes 
fixed shifts by an angle $\Omega$ along the circle's circumference. 
The point is also subject to nonlinear sinusoidal ``corrections,'' 
with the size of the nonlinearity controlled by a parameter $K$. 

The solutions of (\ref{arnold}) are characterized by their winding number
\[\omega=\omega(\Omega,K)=\lim_{n \rightarrow \infty} 
\left[(\theta_n-\theta_0)/n\right],\]  
which can be described roughly as the 
average shift of the point per iteration. 
When the nonlinearity's influence is small, 
this average shift --- and hence the average period --- is determined 
largely by $\Omega$; it may be rational or irrational, with the 
latter being more probable due to the irrationals' pervasiveness. 
As the nonlinearity  $K$ is increased, ``Arnol'd tongues'' --- 
where the winding number $\omega$ locks to a constant rational 
over whole intervals --- form and widen. 
At a critical parameter value, only rational winding numbers are 
left and a complete Devil's Staircase crystallizes. 
Beyond this value, chaos reigns as the system jumps irregularly 
between resonances \cite{Jen84,Sch88}. 

The average cycle length $\bar{P}$ defined for our ENSO system 
of BDEs is clearly analogous to the circle map's winding number,
in both its definition and behavior. 
Note that the QP route to chaos depends in an essential way
on two parameters: $\Omega$ and $K$ for the circle map
and $\beta$ and $\tau$ in our BDE model.

\subsection{ The ``fractal sunburst'': A ``bizarre'' attractor}
\label{sunburst}  
As the system undergoes the transition from an averaged period of 
two to three years a much more complex, and heretofore unsuspected, 
``fractal-sunburst'' structure emerges (Fig.~\ref{fig_burst}, and
first inset in Fig.~\ref{fig_devil2d}).
 As the wave delay $\tau$ is increased, mini-ladders build up, 
collapse or descend only to start climbing up again. 
In the vicinity of a critical value ($\tau \cong 0.5$ years)
that constitutes 
the pattern's focal point, these mini-ladders rapidly condense 
and the structure becomes self-similar, as  each zoom reveals 
the pattern being repeated on a smaller scale.  
We call this a ``bizarre'' attractor because it is more
than ``strange'': strange attractors occur in a system's
phase space, for fixed parameter values, while this
fractal sunburst appears in our model's phase--parameter
space, like the Devil's Staircase. 
The structure in Fig.~4 is attracting, though, only in 
phase space, for fixed parameter values; 
it is, therefore, a generalized attractor, 
and not just a bizarre one.

The influence of the local-process delay $\beta$, along 
with that of the wave-dynamics delay $\tau$, is shown in 
the three-dimensional ``Devil's bleachers'' 
(or ``Devil's terrace,'' according to Jin {\it et al.} \cite{JNG96}) 
of Fig.~\ref{fig_devil3d}.
Note that the Jin {\it et al.} \cite{JNG94,JNG96}
model is an {\em intermediate} model, in the terminology
of modeling hierarchies \cite{GR00}, {\it i.e.}
intermediate between the simplest ``toy models''
(BDEs or ODEs) and highly detailed models based on
discretized systems of PDEs in three space dimensions,
such as the general circulation models (GCMs) used in
climate simulation.
Specifically, the intermediate model of Jin and 
colleagues is based on a system of nonlinear PDEs
in one space dimension (longitude along the equator).
The Devil's bleachers in our BDE model resemble fairly
well those in the intermediate ENSO model of Jin {\it et al.}
\cite{JNG96}. 
The latter, though, did not exhibit a fractal sunburst,   
which appears, on the whole, to be an entirely new addition
to the catalog of fractals \cite{Mandelbrot,PR86,PJS92}.

It would be interesting to find out whether such a bizarre 
attractor occurs in other types of dynamical systems.
Its specific significance in the ENSO problem might be associated 
with the fact that a broad peak with a period 
between two and three years appears in many spectral analyses
of SSTs and surface winds from the Tropical Pacific
\cite{RWR90,JNG95}.
Various characteristics of the Devil's Staircase have been well
documented in both observations \cite{JNG95,MVG98,YSG}
and GCM simulations \cite{GR00,JNG94} of ENSO.
It remains to see whether this will be the case for the 
fractal sunburst as well.

\section{A BDE Model for Seismicity} 
\label{CCM}

Lattice models of systems of interacting elements 
are widely applied for modeling seismicity, 
starting from the pioneering works of 
Burridge and Knopoff \cite{BK67},
All\`{e}gre {\it et al.} \cite{ALP82}, and 
Bak {\it et al.} \cite{BTW88}. 
The state of the art is summarized in 
\cite{KB02,KS99,NGT94,RTK00,TNG00}. 
Recently, colliding-cascade models
\cite{ZKG03a,ZKG03b,GKZ+00a,GKZ+00b} have been able to 
reproduce a wide set of observed characteristics of 
earthquake dynamics \cite{Kei96b,Sch90,Tur97}:
(i) the seismic cycle; 
(ii) intermittency in the seismic regime; 
(iii) the size distribution of earthquakes, known as the  
Gutenberg-Richter relation; 
(iv) clustering of earthquakes in space and time; 
(v) long-range correlations in earthquake occurrence; and 
(vi) a variety of seismicity patterns premonitory to a strong 
earthquake.

Introducing the BDE concept into the modeling of colliding cascades, 
we replace the elementary interactions of elements in the system 
by their integral effect, represented by the delayed switching 
between the distinct states of each element: 
unloaded or loaded, and intact or failed. 
In this way, we bypass the necessity of reconstructing the 
global behavior of the system from the numerous complex and diverse 
interactions that researchers are only mastering by and by and 
never completely.
Zaliapin {\it et al.} \cite{ZKG03a,ZKG03b} have shown that this 
modeling framework does simplify the detailed study of the 
system's dynamics, while still capturing its essential features.
Moreover, the BDE results provide additional insight into
the system's range of possible behavior, as well as into its
predictability. 

\subsection{Conceptual ingredients}
Colliding-cascade models \cite{ZKG03a,ZKG03b,GKZ+00a,GKZ+00b} 
synthesize three processes that play an important role in 
lithosphere dynamics, as well as in many other complex systems:
(i) the system has a hierarchical structure;
(ii) the system is continuously loaded (or driven) 
by external sources; and 
(iii) the elements of the system fail (break down) under the 
load, causing redistribution of the load and strength throughout 
the system. 
Eventually the failed elements heal, thereby ensuring the 
continuous operation of the system.

The load is applied at the top of the hierarchy and transferred 
downwards, thus forming a {\it direct cascade of loading}. 
Failures are initiated at the lowest level of the hierarchy, 
and gradually propagate upwards, thereby forming an {\it inverse 
cascade of failures}, which is followed by healing. 
The interaction of direct and inverse cascades establishes 
the dynamics of the system: loading triggers the failures, 
and failures redistribute and release the load.
In its applications to seismicity, the model's hierarchical structure 
represents a fault network, loading imitates the effect of tectonic forces, 
and failures imitate earthquakes.
 
\subsection{Model structure and parameters}
(i) The model acts on a directed graph whose nodes, except
the top one and the bottom ones, have connectivity six.
Each node, except the bottom ones, is a parent to three children, 
that are siblings to each other. 
This graph is obtained from a directed ternary tree,
which has its root in the top element, by
connecting siblings, {\it i.e.}, groups of three nodes that 
have the same parent.

(ii) Each element possesses a certain degree of {\it weakness} 
or {\it fatigue}. An element fails when its weakness exceeds a certain 
threshold.

(iii) The model runs in discrete time $t=0,1,\dots$. 
At each epoch a given element may be either {\it intact} 
or {\it failed (broken)}, and either {\it loaded} or {\it unloaded}. 
The state of an element $e$ at a epoch $t$ is defined by two 
Boolean functions: 
$s_e(t)=0$, if an element is intact, and
$s_e(t)=1$, if an element is failed;
$l_e(t)=0$, if an element is unloaded, and
$l_e(t)=1$, if an element is loaded.

(iv) An element of the system may switch from one state 
$(s,l)\in\{0,1\}^2$ to another under an impact from its nearest 
neighbors and external sources. 
The dynamics of the system is controlled by the time delays 
between the given impact and switching to another state. 

(v) At the start, $t = 0$, all elements are in the state $(0,0)$, 
intact and unloaded. 
Most of the changes in the state of an element occur in the 
following cycle: 
\[(0,0)\rightarrow(0,1)\rightarrow(1,1)\rightarrow(1,0)\rightarrow(0,0)\dots\]
Other sequences, however, are also possible, except that 
a failed and loaded element may switch only to a failed and 
unloaded state, 
$(1,1)\rightarrow(1,0)$. The latter transition
mimics fast stress drop after a failure. 

(vi) All the interactions take finite, nonzero time. 
We model this by introducing four basic time delays:
$\Delta_L$, between being impacted by the load and switching 
to the loaded state, 
$(\cdot,0)\rightarrow(\cdot,1)$;
$\Delta_F$, between the increase in weakness and switching 
to the failed state, 
$(0,\cdot)\rightarrow(1,\cdot)$;
$\Delta_D$, between failure and switching to the unloaded state, 
$(\cdot,1)\rightarrow(\cdot,0)$; and
$\Delta_H$, between the moment when healing conditions are established 
and switching to the intact (healed) state, $(1,\cdot)\rightarrow(0,\cdot)$.

The duration of each particular delay, from one switch of an element's 
state to the next, is determined from these basic delays, depending on 
the state of the element as well as of its nearest neighbors during the 
preceding time interval (see \cite{ZKG03a} for details). This represents 
yet another generalization of the set of deterministic, autonomous 
equations (\ref{BDE}) with fixed delays $\theta_{ij}$: here the effective 
delays are both variable and state-dependent.

(vii) Failures are initiated randomly within the elements at the lowest
level.
 
The two primary delays in this system are the loading time $\Delta_L$ 
necessary for an unloaded element to become loaded under the impact of its 
parent, and the healing time $\Delta_H$ necessary for a broken
element to recover.

{\bf Conservation law.} 
The model is forced and dissipative, if we associate the loading with an 
energy influx. The energy dissipates only at the lowest level, where it is 
transferred downwards, out of the model. In any part of the model not 
including the lowest level, energy conservation holds, 
but only after averaging over sufficiently large time intervals. 
On small intervals it may not hold, due to the discrete time delays 
involved in energy transfer.

{\bf Model solutions.}
The output of the model is a catalog ${\mathcal{C}}$ of earthquakes ---
{\it i.e.,} of failures of its elements --- similar to 
the simplest routine catalogs of observed earthquakes:
\begin{equation}
\label{sequence}
{\mathcal{C}}=(t_k,~m_k,~h_k),~k=1,2,\dots;~t_{k}\le t_{k+1}.
\end{equation}
In real-life catalogs, $t_k$ is the starting time of the rupture; 
$m_k$ is the magnitude, 
a logarithmic measure of energy released by the earthquake; 
and $h_k$ is the vector that comprises the coordinates of the 
hypocenter. 
The latter is a point approximation of the area 
where the rupture started.
In our BDE model, earthquakes correspond to failed elements, 
$m_k$ is the level at which the failed element is situated
within the directed graph,
while the position of this element within its level is a
counterpart of $h_k$. 

\subsection{Seismic regimes}
A long-term pattern of seismicity within a given region is usually 
called a {\it seismic regime}. 
It is characterized by the frequency and irregularity of the strong 
earthquakes' occurrence, more specifically by 
(i) the Gutenberg-Richter relation, {\it i.e.} the time-and-space 
averaged magnitude--frequency distribution; 
(ii) the variability of this relation with time; and 
(iii) the maximal possible magnitude. 
The notion of seismic regime here is a much more complete 
description of seismic activity than the ``level of seismicity,'' 
often used to discriminate among regions with high, medium, low 
and negligible seismicity; the latter are called aseismic regions.

\begin{figure}[ht]
\centering\includegraphics[width=.7\textwidth]{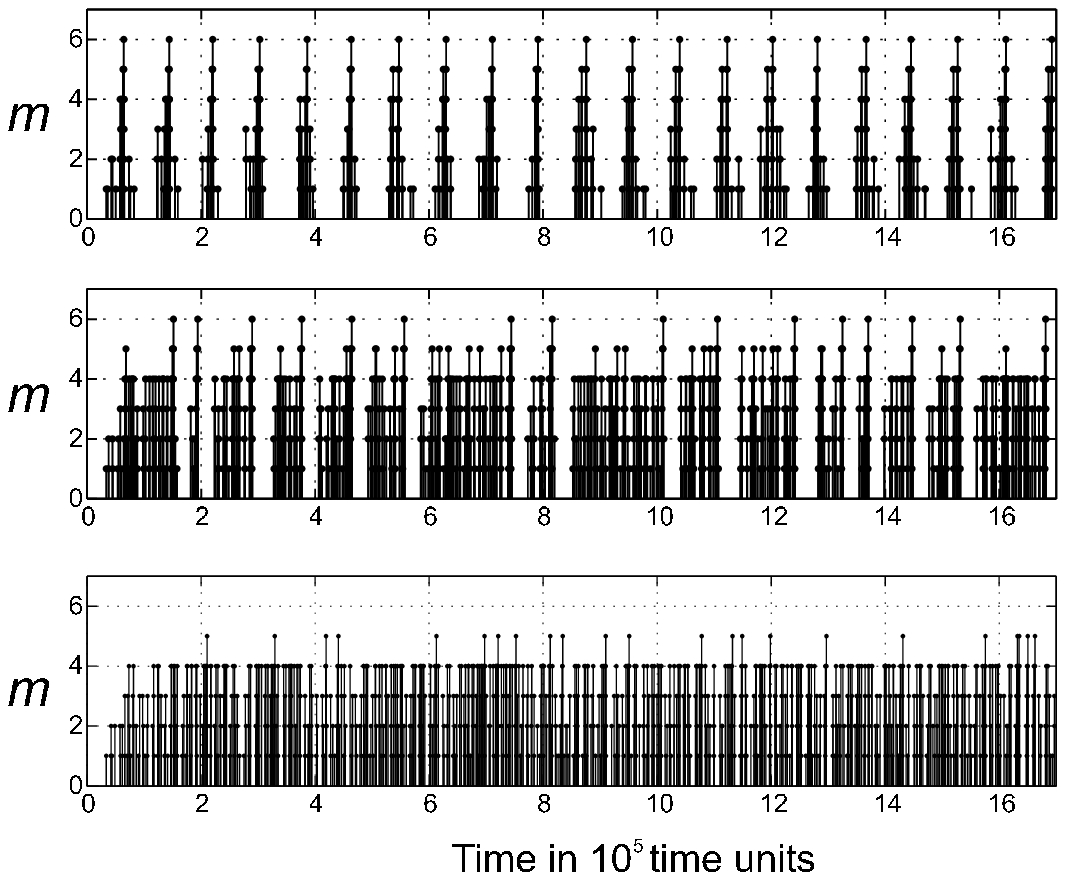}
\caption{Three seismic regimes in a colliding-cascade model of
lithospheric dynamics; each earthquake sequence illustrates
characteristic features of the corresponding regime and only
a small fraction of each sequence is shown.
Top panel -- regime {\bf H} (High), $\Delta_H=0.5\cdot10^4$; 
middle panel -- regime {\bf I} (Intermittent), $\Delta_H=10^3$;
bottom panel -- regime {\bf L} (Low), $\Delta_H=0.5\cdot10^3$;
in all three panels $\Delta_L=0.5\cdot10^4$ (see also Fig.~\ref{fig_regimes}
below) and  
the number of nodes in the simulated lattice is $1093$, for 
a tree depth of $L=6$,
the maximum magnitude of any earthquake in the BDE model.
Reproduced from Zaliapin {\it et al.} \cite{ZKG03a}, with kind 
permission of Springer Science and Business Media. 
} 
\label{fig_reg3} 
\end{figure}

\begin{figure}[ht]
\centering\includegraphics[width=.7\textwidth]{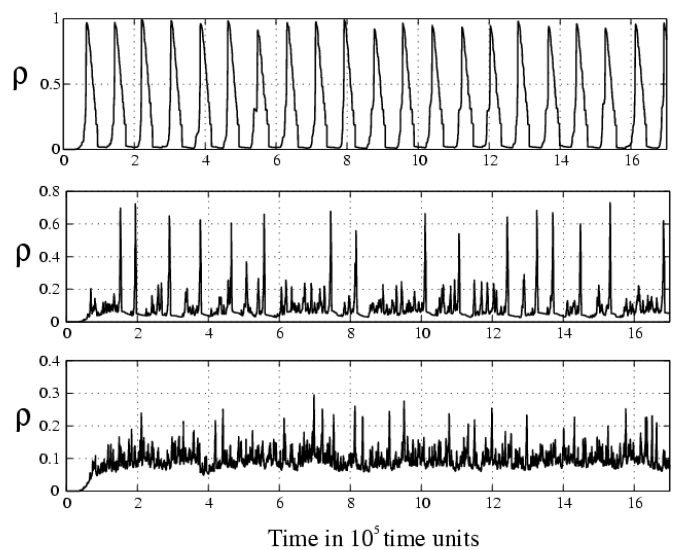}
\caption{Three seismic regimes: internal dynamics of the BDE model. 
The panels show the density $\rho(n)$ of broken elements in the system, 
as defined by Eq.~(\ref{rho});
they correspond to the synthetic sequences 
shown in Fig.~\ref{fig_reg3}.
Top panel -- Regime {\bf H}; 
middle panel -- Regime {\bf I}; and 
bottom panel -- regime {\bf L}.
Reproduced from Zaliapin {\it et al.} \cite{ZKG03a}, with kind 
permission of Springer Science and Business Media. 
} 
\label{fig_rho3} 
\end{figure}

\begin{figure}[ht]
\centering\includegraphics[width=.7\textwidth]{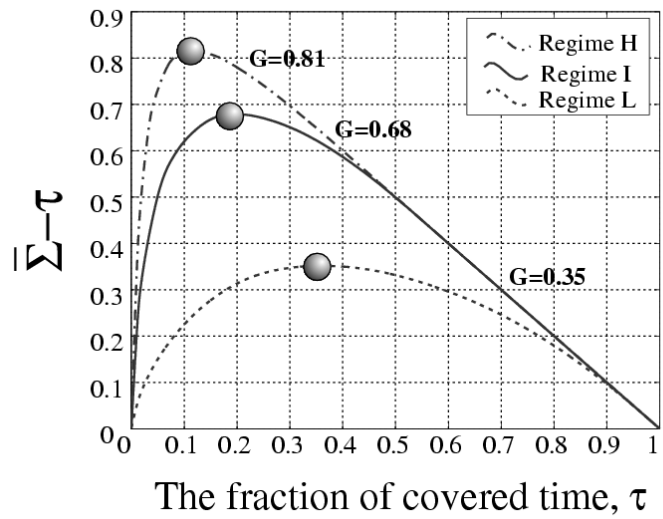}
\caption{Measure $\mathcal{G}(I)$ of seismic clustering
in our BDE model of colliding cascades; 
see Eq. (\ref{G}).
The three curves correspond to the three synthetic sequences 
shown in Fig.~\ref{fig_reg3}.
Reproduced from Zaliapin {\it et al.} \cite{ZKG03a}, with kind 
permission of Springer Science and Business Media.}
\label{fig_G} 
\end{figure}

\begin{figure}[ht]
\centering\includegraphics[width=.7\textwidth]{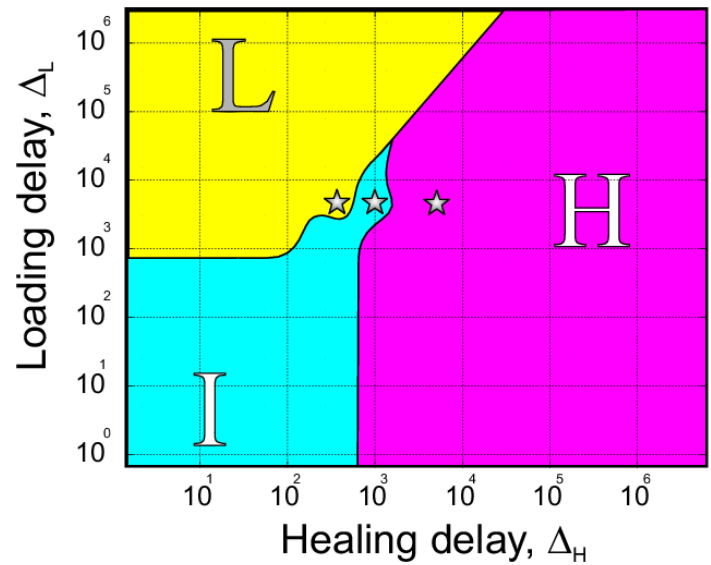}
\caption{Regime diagram in the $(\Delta_L,\Delta_H)$ plane of
the loading and healing delays.
Stars correspond to the sequences shown in Fig.~\ref{fig_reg3}.
Reproduced from Zaliapin {\it et al.} \cite{ZKG03a}, with kind 
permission of Springer Science and Business Media.} 
\label{fig_regimes} 
\end{figure}

The seismic regime is to a large extent determined by the neotectonics 
of a region; this involves, roughly speaking, two factors: 
(i) the rate of crustal deformations; and 
(ii) the crustal consolidation, determining what part of deformations 
is realized through the earthquakes. 
However, as is typical for complex processes, the long-term patterns 
of seismicity may switch from one to another in the same region, as well as 
migrate from one area to another on a regional or global 
scale \cite{PA95,Rom93}. 
Our BDE model produces synthetic sequences that can be divided into 
three seismic regimes, illustrated in Figs.~\ref{fig_reg3}--\ref{fig_twoD}.

Regime {\bf H}: {\it High and nearly periodic seismicity} 
(top panel of Figs.~\ref{fig_reg3} and \ref{fig_rho3}). 
The fractures within each cycle reach the top level, $m = L$, where
our underlying ternary graph has depth $L=6$.
The sequence is approximately periodic,
in the statistical sense of cyclo-stationarity \cite{Mer99}.

Regime {\bf I}: {\it Intermittent seismicity} 
(middle panel of Figs.~\ref{fig_reg3} and \ref{fig_rho3}). 
The seismicity reaches the top level for some but not all cycles,
and cycle length is very irregular.

Regime {\bf L}: {\it Medium or low seismicity} 
(lower panel of Figs.~\ref{fig_reg3} and \ref{fig_rho3}). 
No cycle reaches the top level and seismic activity is much more constant
at a low or medium level, without the long quiescent intervals present
in Regimes {\bf H} and {\bf I}.

The location of these three regimes in the plane of the two 
key parameters $(\Delta_L, \Delta_H)$ is shown in Fig.~\ref{fig_regimes}.
Figures \ref{fig_reg3}--\ref{fig_tr} were computed for a
tree depth of $L=6$, {\em i.e.} 1093 nodes.
Many calculations were also carried out for $L=7$, {\em i.e.} 3280 nodes, 
and the results were similar, but are not reported here.

\subsection{Quantitative analysis of regimes} 
The quantitative analysis of model earthquake sequences
and regimes is facilitated by the two measures described
below.

{\bf Density of failed elements.}
The density $\rho(t)$ of the elements that are 
in a failed state at the epoch $n$ is given by:
\begin{equation}
\label{rho}
\rho(t)=\left[\nu_1(t)+\dots+\nu_L(t)\right]/L.
\end{equation}
Here $\nu_m(t)$ is the fraction of failed elements at 
the $m$-th level of the hierarchy at the epoch $t$, 
while $L$ is the depth of the underlying tree.
Sometimes we consider this measure
averaged over a time interval, or a union of intervals, $I$ and denote 
it by $\rho(I)$. 
The density $\rho(t)$ for the three sequences of Fig.~\ref{fig_reg3}
is shown in Fig.~\ref{fig_rho3}.

{\bf Irregularity of energy release.}
The second measure is the irregularity $\mathcal{G}(I)$ of energy release
over the time interval $I$.
It is motivated by the fact that one of the major differences between 
regimes resides in the temporal character of seismic 
energy release. 
The measure $\mathcal{G}$ is defined by the following sequence of steps:

(i) First, define a measure $\Sigma(I)$ of seismic 
activity within the time interval, or union of time intervals, 
$I$ as
\begin{equation}
\label{Sigma}
\Sigma(I)=\frac{1}{n_I}\sum_{i=1}^{n_I} 10^{Bm_i},~B=\log_{10}3.
\end{equation}
The summation in (\ref{Sigma}) is taken over all events within 
$I$, {\it i.e.}, $t_i\in I$; $n_I$ is the total number of 
such events, and $m_i$ is the magnitude of the $i$-th event. 
The value of $B$ equalizes, on average,
the contribution of earthquakes with different magnitudes, 
that is from different levels of the hierarchy.
In observed seismicity, $\Sigma(I)$ has a transparent 
physical meaning: 
given an appropriate choice of $B$, it estimates the 
total area of the faults unlocked by the earthquakes 
during the interval $I$ \cite{KM64}.
This measure is successfully used in several earthquake 
prediction algorithms \cite{KB02}.

(ii) Consider a subdivision of the interval $I$ into a
set of nonoverlapping intervals of equal length $\epsilon>0$.
For simplicity we choose $\epsilon$
such that $|I|=\epsilon N_I$, where
$|\cdot|$ denotes the length of an interval and
$N_I$ is an integer.
Therefore, we have the following representation:
\begin{equation}
\label{span}
I=\bigcup_{j=1}^{N_I} I_j, \hspace{.1in}|I_k|=
\epsilon,\hspace{.1in} k=1,\dots,N_I; 
\hspace{.1in} I_j\cap I_k=\emptyset~{\rm for~} j\ne k.
\end{equation}

(iii) For each $k=1,\dots,N_I$ we choose a 
$k$-subset
\[\Omega(k)=\bigcup_{i=i_1,\dots,i_k} I_i\]  
that maximizes the value of the accumulated $\Sigma$:
\begin{equation}
\label{subset}
\Sigma[\Omega(k)]\equiv\Sigma^*(k)=
\max\limits_{(i_1,\dots,i_k)}\left\{\Sigma \left[\bigcup_{j=1}^{k} I_{i_j}
\right]\right\}.
\end{equation}
Here the maximum is taken over all $k$-subsets of the covering
set (\ref{span}).

(iv) Introducing the notations
\begin{equation}
\label{notation}
\bar\Sigma(k)=\Sigma^*(k)/\Sigma(I), ~\tau(k)=k\epsilon/|I|,
\end{equation}
we finally define the measure $\mathcal{G}$
of clustering within the interval $I$ as
\begin{equation}
\label{G}
\mathcal{G}(I)=\max_{k=1,\dots,N_I}\left\{\bar\Sigma(k)-\tau(k)\right\}.
\end{equation}

Figure~\ref{fig_G} illustrates this definition
by displaying the curves $\bar\Sigma-\tau$ vs. $\tau$ 
for the three synthetic sequences shown in Fig.~\ref{fig_reg3}.
The curves give, essentially, the maximum seismic activity
minus the mean activity, as a function of length of time
over which the activity occurs, and
the maximum of each curve gives the corresponding value
of $\mathcal{G}$.
The more clustered the sequence, the more convex is the
corresponding curve, the larger the corresponding
value of $\mathcal{G}$, and the shorter the interval
for which this value of $\mathcal{G}$ is realized.
Despite its somewhat elaborate definition, $\mathcal{G}$ 
has a transparent intuitive interpretation: 
it equals unity for a catalog consisting of a single event 
(delta function, burst of energy), 
and it is zero for a marked Poisson process 
(uniform energy release).
Generally, it takes values between 0 and 1 depending on
the irregularity of the observed energy release.

\subsection{Bifurcation diagram}
\begin{figure}[ht]
\vspace{-.3cm}
\centering\includegraphics[height=.7\textheight]{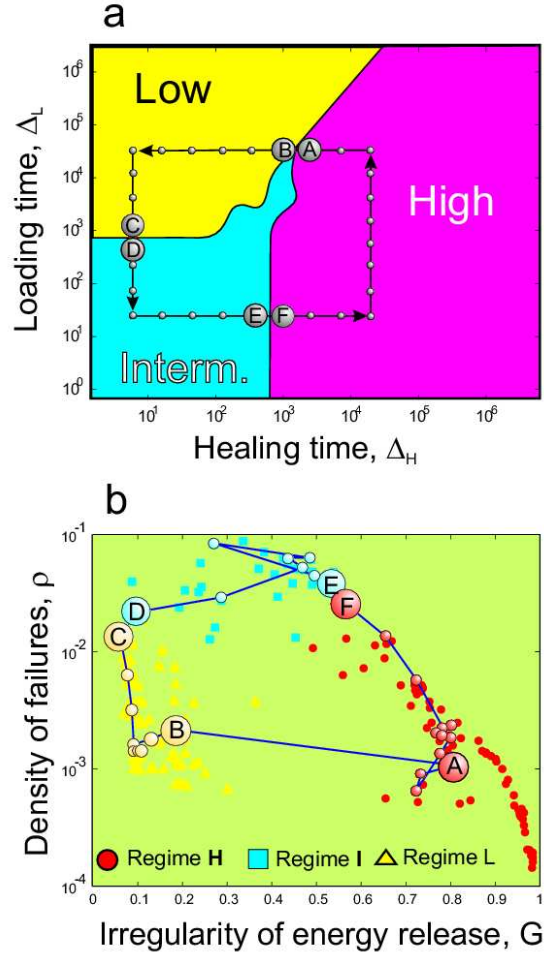}
\vspace{-.3cm}
\caption{Bifurcation diagram for the BDE seismic model:
(a) rectangular path in the delay plane $(\Delta_L,\,\Delta_H)$;
and (b) the measures $\mathcal{G}$ and $\rho$,
calculated along the rectangular path shown in panel (a).
The transition between points (A) and (B), 
{\it i.e.} between regimes {\bf H} and {\bf L}, is very sharp 
with respect to the change in irregularity $\mathcal{G}$ of 
energy release, while almost negligible with respect to the 
change in failure density $\rho$.
The colored circles, triangles, and squares in panel (b)
correspond to synthetic catalogs from regimes {\bf H},
{\bf I}, and {\bf L}, respectively; these catalogs 
are produced for the points indicated along the rectangular path 
in panel (a), as well as for many scatter points that lie on a uniform grid 
covering the entire regime diagram, with the same resolution in
$\delta_H$ and $\delta_L$ as those along the path.}
\label{fig_twoD} 
\end{figure}

\begin{figure}[ht]
\centering\includegraphics[height=.8\textheight]{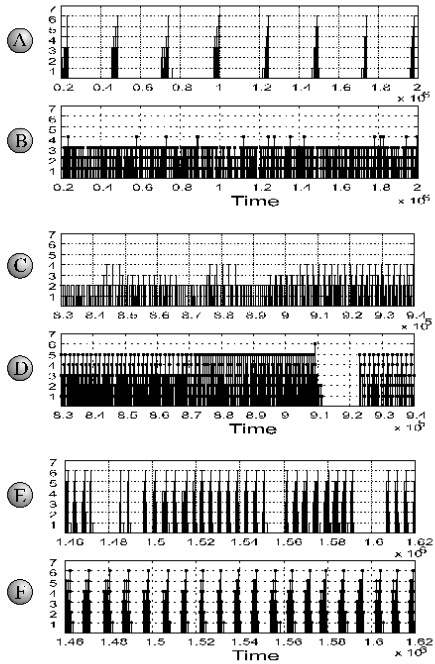}
\caption{Synthetic sequences corresponding to the key points 
along the rectangular path in parameter space of Fig.~\ref{fig_twoD}a.
The panels illustrate the transitions between the regimes 
{\bf H} and {\bf L} --- panels (A) and (B);
{\bf L} and {\bf I} --- (C) and (D); and 
{\bf I} and {\bf H} --- (E) and (F). 
The transition from (A) to (B) is very pronounced,
while the other two transitions are smoother.
Reproduced from Zaliapin {\it et al.} \cite{ZKG03a}, with kind 
permission of Springer Science and Business Media. } 
\label{fig_tr} 
\end{figure}

Figure~\ref{fig_twoD} provides a closer look at the regime 
diagram of Fig.~\ref{fig_regimes}: it illustrates the transition 
between regimes in the parameter plane $(\Delta_L,\Delta_H)$.
To do so, Fig.~\ref{fig_twoD} (a) shows a rectangular path 
in the parameter plane that passes through all three regimes and 
touches the triple point.
We single out 30 points along this path; they
are indicated by small circles in the figure.
The three pairs of points that correspond to the transitions between
regimes are distinguished by larger circles and marked in 
addition by letters, for example (A) and (B) mark the 
transition from Regime {\bf H} to Regime {\bf L}.

We estimate the clustering $\mathcal{G}(I)$ and average density 
$\rho(I)$ over the time interval $I$ of length $2\cdot10^6$ time units,
for representative synthetic sequences that correspond to the 
30 marked points along the rectangular path in
Fig.~\ref{fig_twoD}a.
Figure~\ref{fig_twoD}b is a plot of $\rho(I)$ vs. $\mathcal{G}(I)$
for these 30 sequences.
The values of $\mathcal{G}$ drop dramatically, from 0.8 to 0.18, 
between points (A) and (B): this
means that the energy release switches from highly irregular
to almost uniform between Regimes {\bf H} and {\bf L}.
This transition, however, barely changes the average
density $\rho$ of failures.

The transitions between the other pairs of regimes are 
much smoother. 
The clustering drops further, from $\mathcal{G}=0.18$ to 
$\mathcal{G}\approx 0.1$, and then remains at the latter low level
within Regime {\bf L}.
It increases gradually, albeit not monotonically,
from 0.1 to 0.8 between points (C) and (A), 
on its way through regimes {\bf I} and {\bf H}.
The increase of $\Delta_L$ along the right side of the 
rectangular path in Fig.~\ref{fig_twoD}a, between points
(F) and (A), corresponds to a decrease of 
$\rho$ and a slight increase of clustering $\mathcal{G}$, 
from $0.5$--$0.6$ to $\approx 0.8$.  
 
The transition between regimes is illustrated further  
in Fig.~\ref{fig_tr}.
Each panel shows a fragment of the six synthetic sequences
that correspond to the points (A)--(F) in Fig.~\ref{fig_twoD}a.
The sharp difference in the character of the energy release
at the transition between Regimes {\bf H} (point (A)) 
and {\bf L} (point (B)) is very clear, here too.
The other two transitions, from (C) to (D)
and (E) to (F), are much smoother.
Still, they highlight the intermittent 
character of Regime {\bf I}, to which
points (D) and (E) belong.

Zaliapin {\it et al.} \cite{ZKG03b} considered applications 
of these results to earthquake prediction. 
These authors used the simulated catalogs to study in greater 
detail the performance of pattern recognition methods tested 
already on observed catalogs and other models 
\cite{KB02,KM64,JS99,Kei94,PCS94,MDR+90,KLK+96,BV93,BOS+98}, 
devised new methods, and experimented with combination of different 
individual premonitory patterns into a collective prediction algorithm.

\section{BDEs on a Lattice and Cellular Automata (CAs)} 
\label{PBDE}
While the development and applications of BDEs
started about 25 ago, this is a very short
time span compared to the long history of ODEs, PDEs, maps, and even
CAs.
The BDE results obtained so far, though, are sufficiently
intriguing to warrant further exploration.
In this section, we provide some preliminary results on BDE systems
with a large or infinite number $N$ of variables, and we discuss 
in greater detail their connections with CAs \cite{Wol94,vN,We91}, 
(see also Fig.~1 and Sect. 2.6).

These ``partial BDEs'' that we are led to explore,
with $N \rightarrow \infty $ Boolean
variables, were mentioned in passing in \cite{Mull84}, and stand 
in the same relation to ``ordinary BDEs,'' explored so far, 
as PDEs do to ODEs.
The classification of what we could call now {\it ordinary}
BDEs into conservative and dissipative (Sect.~\ref{BDE_sect})
suggests that {\it partial} BDEs of different types can
exist as well. 
 
\subsection{Towards partial BDEs of hyperbolic and parabolic type}

We want, first of all, to clarify what the ``correct''
BDE equivalent of partial derivatives may be. To do so, we start by studying
possible candidates for {\em hyperbolic} and {\em parabolic} partial BDEs.
Intuitively, these should correspond to generalizations of conservative and 
dissipative BDEs, respectively (see Sects.~2.3 and 2.5), that are 
infinite-dimensional (in phase space). 
We are thus looking for the discrete-variable 
version of the typical hyperbolic and parabolic PDEs:
\begin{eqnarray}
\frac{\partial}{\partial t} {v}(z,t)&=&\frac{\partial}{\partial z} 
{v}(z,t),
\label{hyp}\\
\frac{\partial}{\partial t} {v}(z,t)&=&\frac{\partial^2}{\partial z^2} 
{v}(z,t),
\label{par}
\end{eqnarray}
where, in the spatially one-dimensional case, 
$v: \mathbb{R}^2 \rightarrow \mathbb{R}$. We 
wish to replace the real-valued function $v$ with a Boolean 
function $u: \mathbb{R}^2 \rightarrow \mathbb{B}$, {\it i.e.} $u(z,t)=0,1$, 
with $z\in\mathbb{R}$ and $t\in\mathbb{R^+}$, while 
conserving the same qualitative behavior of the solutions.

Since for Boolean-valued variables 
$|x-y|=x \bigtriangledown y$ (see Sect.~\ref{BDE_sect} and Table~1), 
one is tempted to use the
``eXclusive OR'' (XOR) operator $\bigtriangledown$ 
for evaluating differences. Moreover, one is led to introduce
a $time$ delay $\theta_t$ and a $space$ delay $\theta_z$
when approximating the derivatives in
Eqs.~(\ref{hyp}) and (\ref{par}) by finite differences. First-order 
expansions then lead to the equations:
\begin{equation}
u(z,t+\theta_t)\bigtriangledown u(z,t)=
u(z+\theta_z,t)\bigtriangledown u(z,t)
\label{hypbde}
\end{equation}
in the hyperbolic case, and
\begin{equation}
u(z,t+\theta_t)\bigtriangledown u(z,t)=
u(z-\theta_z,t)\bigtriangledown u(z+\theta_z,t)
\label{parbde}
\end{equation}
in the parabolic one. 

\subsection{Boundary conditions and discretizing space}
\label{discr}
The pure Cauchy problem for Eq.~\eqref{hypbde} 
on the entire real line \cite{CH62}, $z \in (-\infty,\infty)$
has solutions of the form:
\begin{equation}
u(z,t)=u(z+\theta_z,t-\theta_t). 
\label{hyper} 
\end{equation}
This first step towards a partial BDE equivalent of a hyperbolic PDE
displays therewith the expected behavior of a ``wave'' propagating in the 
$(z,t)$ plane.
The propagation is from right to left for
increasing times when the ``plus'' sign is chosen in the 
right-hand side (rhs) of Eq.~(\ref{hypbde}), 
as we did, but it  could be in the opposite direction choosing the 
``minus'' sign instead. 
The solution \eqref{hyper} of \eqref{hypbde} 
exists for all times $t \ge \theta_t$ and it is unique for all delays 
$(\theta_z, \theta_t) 
\in (0,1]^2$, and for all initial data 
$u_0(z,t)$ with $z \in (-\infty,\infty)$ and $t \in [0,\theta_t)$.

In continuous space and time, Eq.~\eqref{hypbde} under consideration is 
conservative and invertible. Aside from the pure Cauchy problem discussed
above, when $u_0(z,t)$ is given for $z \in (-\infty,\infty)$, one can also
formulate for \eqref{hypbde} a space-periodic initial boundary value problem 
(IBVP), with $u_0(z,t)$ given 
for $z \in [0,T_z)$ and $u_0(z+ T_z,t)=u_0(z,t)$. 
The solution of this IBVP displays periodicity in time as well, 
with $T_t=T_z \theta_t / \theta_z$. This time-and-space periodic solution
exists for all time and is unique under conditions that are analogous
to those stated for the pure Cauchy problem above.

Next, we analyze a discrete version of Eq.~(\ref{hypbde}),
which is obtained by studying the evolution of the system on 
a 1-D lattice. Specifically, one considers the grid 
$\{ z_i = z_0+i\theta_z\}_{i \in \mathbb{Z}}$ for a fixed 
$z_0 \in \mathbb{R}$ and assumes that the initial state is constant within the 
space intervals $I_i=[z_0+i\theta_z, z_0+(i+1) \theta_z]$ for 
$t \in [0,\theta_t)$, {\em i.e.},
$u_0(z,t)= \sum_{i=-\infty}^{\infty} u_{0,i}(t){\bf I}_{i}$,
where ${\bf I}_i$ is the characteristic function of the interval
$I_i$, equal to unity on $I_i$ and to zero outside it.
Correspondingly, the behavior
of $u(z,t)$ is determined by the evolution of the elements of the set 
$\{ u_i(t)  \equiv u(z_i,t)  \}_{i \in \mathbb{Z}}$ and 
one gets:
\begin{equation}
u_i(t)=u_{i-1}(t-\theta_t), \hspace{.2in} i\in \mathbb{Z}.
\label{hypbdediscr}
\end{equation}
Equation \eqref{hypbdediscr} is a simple linear BDE system \eqref{pl}, with 
$c_{ij}=\delta_{j,i-1}$ and equal delays $\theta_{ij}=\theta_t$ for all 
$i$ and $j$, in the limit $n \rightarrow \infty$. 
The discretization thus makes it more evident that, in the IBVP case, where 
$\{ u_{0,i}(t) = u_{0,i+N}(t) \}_{i \in \mathbb{Z}}$, 
the solution is immediately periodic in time, without transients, and
with period $T_t=N \theta_t$. 
Notice that for all $(\theta_z, \theta_t) \in (0,1]^2$, whether rational or 
not, one can choose the space discretization generated by multiples 
of $\theta_z$
and the resulting partial BDE system with constant initial data
within the space intervals $I_i$ will depend on the
single delay $\theta_t$. The same observation 
should apply to more general cases, and therefore partial BDEs,
whether hyperbolic or not, but
containing only one space delay $\theta_z$ and one time delay $\theta_t$
cannot display solutions with increasing complexity.
The case of partial BDEs with higher time derivatives and
admitting therewith more than one time delay, is left for future work.

We merely note here that approximating $\partial_{z}$ in
the hyperbolic PDE \eqref{hyp} to the second order 
\cite{RiMo} yields the same partial BDE \eqref{parbde} 
that was obtained from the first-order approximation of the parabolic PDE:
\begin{equation}
u(z,t)=
u(z-\theta_z,t-\theta_t)\bigtriangledown u(z,t-\theta_t) 
\bigtriangledown u(z+\theta_z,t-\theta_t).
\label{hyperbde2}
\end{equation}
The equivalence is apparent by using the associativity of addition
(mod 2) in $\mathbb{B}$ (see Sect.~\ref{BDE_sect}).

This result is slightly, but not utterly unexpected: it is fairly
well known in the numerical analysis of PDEs (i) that higher-order
finite-difference approximations of partial derivatives can lead
to P$\triangle$Es (see Fig.~1) that are consistent with a different
PDE, containing higher-order derivatives \cite{VL79}; and (ii)
that such higher-order approximations, when they are stable,
are more likely than not to be dissipative. Still, Eqs.~\eqref{hypbde},
\eqref{parbde} and \eqref{hyperbde2} show that finding the ``correct'' partial
BDE equivalent of a given PDE is not quite trivial. 
To get further insight into the behavior of Eq.~(\ref{hyperbde2}),
or, equivalently, Eq.~\eqref{parbde}, we 
consider again the case of discrete space, and show that this system
is equivalent, in turn, to a particular CA in the limit of 
infinite size.

\subsection{Partial BDE \eqref{hyperbde2} and ECA rule 150}
\label{rule}
Applying the space discretization scheme from the previous subsection 
(Sect.~\ref{discr}) and using the same space-periodic initial data, 
one finds that
\begin{equation}
u_i(t)=
u_{i-1}(t-\theta_t)\bigtriangledown u_i(t-\theta_t) 
\bigtriangledown u_{i+1}(t-\theta_t), 
\hspace{.2in} i \in \mathbb{Z}.
\label{eca150}
\end{equation}
In order to establish the equivalence of Eq.~\eqref{eca150} with the general
form (\ref{ECA}) of ECA evolution (in one space dimension),
one needs to specify the boundary conditions. For periodic boundary conditions 
$u_{i}=u_{i+N}$, and the number $n$ of BDEs that we can associate with 
such an ECA is $n=2N+1$; this number, though, can be slightly lower or higher 
if one uses Dirichlet or Neumann boundary conditions. 

To identify the particular Boolean function that gives
the ECA corresponding to Eq.~(\ref{eca150}), recall that in an ECA, {\em i.e.}
a 1-D CA where the interactions involve only the nearest right 
and left neighbors $u_{i\pm 1}$, the single rule valid at all sites can be 
described by a binary string. This string summarizes the truth table of the 
rule, by assigning the values of the inputs $(u_{i-1},u_i,u_{i+1})$ in 
decreasing order, from 111 to 000. Correspondingly, one gets $2^3=8$ binary 
digits, 1 or 0, for each possible output. 
The 8-digit string that characterizes the 3-site rule on the rhs of 
Eq.~(\ref{eca150}) is 10010110. 
For brevity, Wolfram replaces \cite{Wol94} each such 8-digit binary number by 
its decimal representation, which yields, in this case, rule 150 \cite{Wol83}. 
We also recall that, for a finite number of variables, {\it i.e.} 
for $i \in  [-N,N]$ with finite size $N$, the ECA version of our 
pigeon-hole lemma (Theorem~2.3 in Sect.~2.2) states that all solutions 
of such an automaton have to become stationary or purely periodic after
a finite number of time steps. 
For $N$ infinite, however, rule 150 yields interesting behavior, with 
self-similar, fractal patterns embedded in its spatio-temporal structure. 
 
Next, we study in detail the simple case of Eq.~(\ref{eca150}) with
the time-constant initial state, $u_{0,i}(t)\equiv u_i(0)$ for 
$0 \le t <\theta_t$, and assume $\theta_t=1$ without loss of generality,
and $\theta_z=\theta_t$ for simplicity. We thus verify merely that our 
partial BDE does generate the rule-150 ECA behavior, 
which is already quite interesting, while we expect less trivial 
BDEs to yield distinct, and possibly even more interesting, behavior.

\begin{figure}[ht]
\centering
\includegraphics[width=.3\textwidth,angle=270]{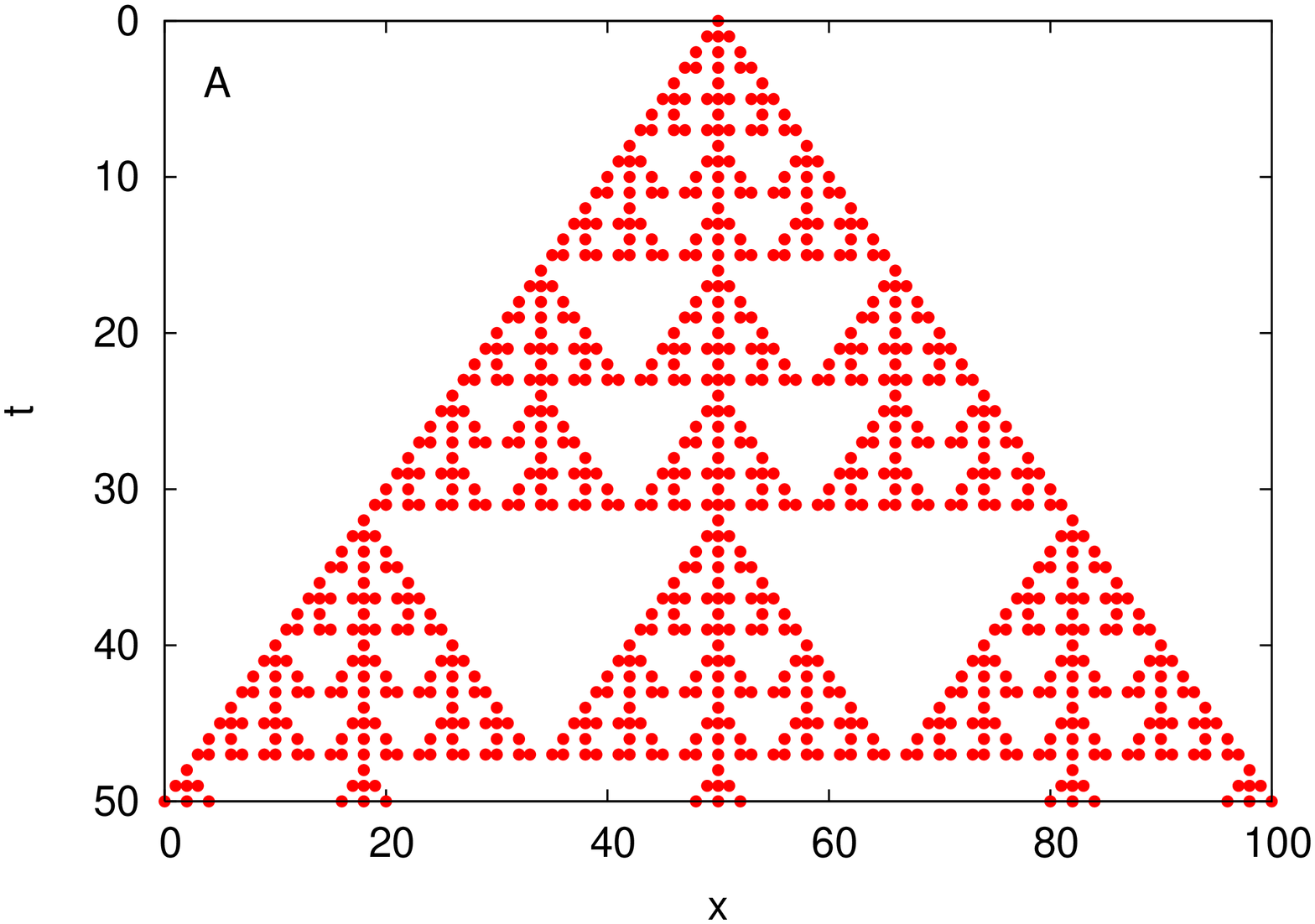}
\includegraphics[width=.3\textwidth,angle=270]{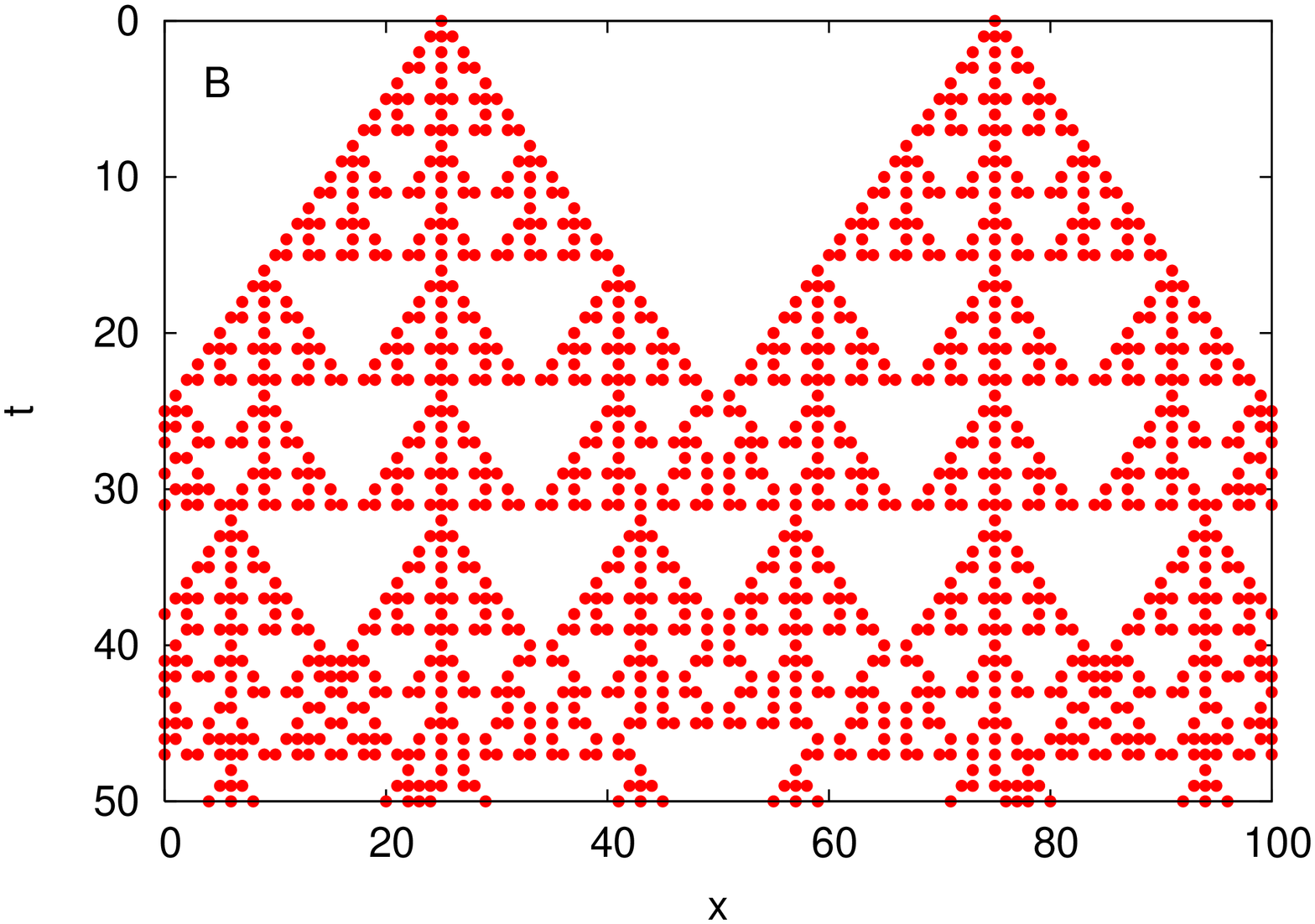}
\caption{Solutions of the ``partial BDE'' (\ref{eca150}): 
(a) for a single nonzero site at $t=0$; and 
(b) the collision of two ``waves,'' each originating from 
such a site.
For the space and time steps $\theta_z=\theta_t=1$, this BDE is equivalent to
the elementary cellular automaton (ECA) with rule 150;
empty sites ($u_i(j)=0$) in white and occupied sites ($u_i(j)=1$) in black. 
} 
\label{fig13} 
\end{figure}

We notice that the system's dynamics for any initial state 
of the pure Cauchy problem can be obtained 
from the evolution of the corresponding ECA that starts from initial data
with a single nonzero value, which we show in Fig.~\ref{fig13}a. 
This property of solutions is due to the fact that rule 150 exhibits the 
important simplifying feature of ``additive superposition'' \cite{Wol83}, 
as evident from Fig.~\ref{fig13}b, where the collision of two ``waves'' is 
plotted; this feature is the result of linearity (mod 2), as discussed
in \cite{Mull84,GhilMull85} and in Sect.~\ref{BDE_sect} here.

In the IBVP case, the 
solutions can behave quite differently, depending on the value of $N$ and on 
the initial state.
We report here only the results for the IBVP with periodic initial data
$u_{i}(0)=u_{i+T_z}(0)$ and number of variables $2N+1$, with $N$ chosen to be 
an integer multiple of the $space$ period $T_z$. 
In this case --- apart from 
the presence of possible fixed points, in particular for $T_z \le 3$ --- the 
solutions of Eq.~(\ref{eca150}) display the longest $time$ periods when 
$T_z=N$. More precisely, $u_0(0)=u_{T_z}(0)=1$ and $u_i(0)=0$ for $i$ 
not a multiple of $T_z$; this choice of the initial state has the 
longest-possible $space$ period at fixed size $N$. 

In agreement with other known results for ECA 
150 \cite{Wol83,MaOdWol84}, we find that our solutions with $T_z=N$ are 
immediately $time$ periodic for $N$ not a multiple of three, whereas there is 
a transient  when $N=3p, p\in\mathbb{N}$; this transient is of length 1 for 
$N$ odd and of length $2^{j-1}$, where $2^j$ is the largest power of two which 
divides $N$, otherwise. The ECA with rule 150 belongs  to the third class in 
Wolfram's classification \cite{Wol83,Wol84}, in which evolution for 
$N\to\infty$ can lead to aperiodic, chaotic patterns. While the behavior
of this ECA is predictable from the knowledge of the initial state
for the IBVP with finite $N$, the length of the time period $T_z$ can 
increase rapidly with $N$. In particular, one finds a time period 
already as long as $511$ for space-periodic initial data with 
$T_z=N=19$. 

\begin{figure}[hpbt]
\centering
\includegraphics[width=.6\textwidth,angle=270]{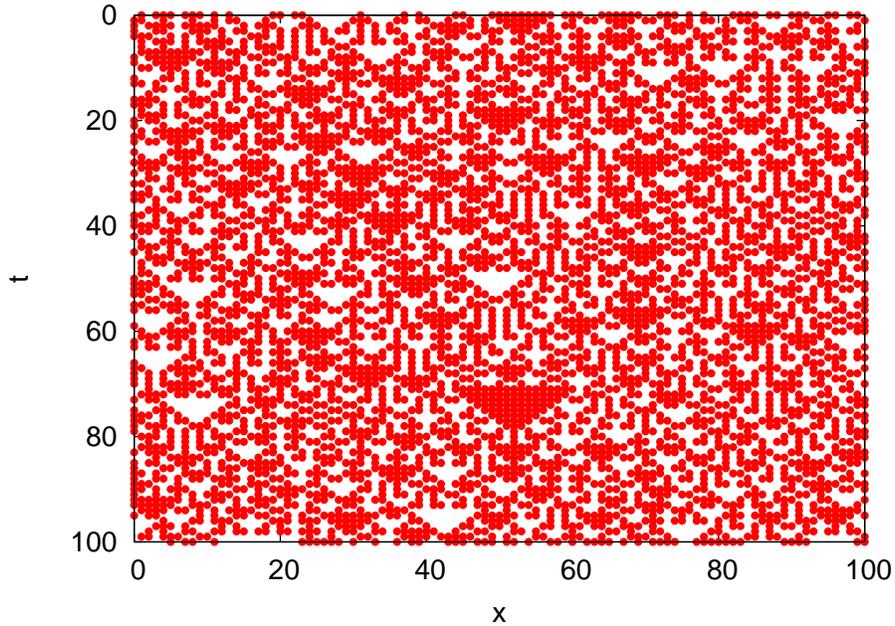}
\caption{
The solution of the BDE (\ref{eca150}) starting from a random initial state of 
length $N=100$. The qualitative behavior is characterized by ``triangles'' of 
empty ($u_i(j)=0$) or occupied ($u_i(j)=1$) sites but without any recurrent 
pattern; this behavior does not depend on the particular random initial state.}
\label{fig16}
\end{figure}

To show that the behavior of the IBVP for Eq.~\eqref{hyperbde2} is not 
periodic in time in the 
limit of $N \rightarrow \infty$, we present in Fig.~\ref{fig16} the results 
for a random initial state of length $N=100$. These 
results do not show any ``recurrent pattern,'' apart from the expected 
\cite{Wol94,We91} appearance of characteristic ``triangles''
that still emerge from the chaotic distribution of empty and occupied
sites.

We have seen in Sect.~2 (see Definition 2.1) that 
ordinary BDEs are conservative if their behavior is immediately periodic 
for any initial data, and that correspondingly they are also time-reversible. 
Our findings show that Eq.~(\ref{hyperbde2}) is dissipative, since one 
observes transients for the periodic IBVP with various $N$ values;
this is expected from its equivalence with the ``parabolic'' BDE
\eqref{parbde}. 
Nevertheless, in the ordinary BDE context, the connectives of the 
form (\ref{hyperbde2}), based on the $\bigtriangledown$ operator, are often 
conservative at least for some set of 
distinct delays. We wish therefore to extend the analysis 
of this partial BDE ---from the case of all delays being equal,
$\theta_{ij}\equiv \theta_t$, as above--- to the case of distinct 
$\theta_t$ values in the rhs of Eq.~(\ref{eca150}), since this could lead
to conservative behavior and to increasingly complex solutions
for some set of delays. Doing so, however, goes beyond the purpose
of the present paper and will have to await future work.

Another attempt at formulating parabolic partial BDEs is given
by the equations:
\begin{eqnarray}
u(z,t)&=&[u(z-\theta_z,t-\theta_t) \lor u(z+\theta_z,t-\theta_t)] 
\bigtriangledown u(z,t-\theta_t), \label{dissor} \\
u(z,t)&=&[u(z-\theta_z,t-\theta_t) \wedge u(z+\theta_z,t-\theta_t)] 
\bigtriangledown u(z,t-\theta_t), \label{dissand} 
\end{eqnarray}
obtained from the parabolic PDE \eqref{par} by replacing $\partial_{zz}$ 
with the OR and the AND operator, respectively. From the known results on 
ordinary BDEs, the solutions in these cases can be expected to reproduce 
closely the behavior of dissipative PDEs. This intuitive conjecture is 
confirmed by the analysis in terms of their ECA equivalent. Thus, 
Eq.~(\ref{dissor}) corresponds to rule 54, in the first class, and 
Eq.~(\ref{dissand}) to rule 108, in the second class. Accordingly,
the evolution of their solutions leads to fixed points or to small-period 
limit cycles, respectively; in neither case does the $N \rightarrow \infty$ 
limit display chaotic behavior. We thus expect these connectives to be good 
candidates as simple examples of ``correct'' partial BDE equivalents of 
dissipative PDEs. 

\subsection{Summary on partial BDEs and future work}
Our results on classifying finite BDE systems, on the one hand, and our
replacing partial derivatives in PDEs by Boolean operators,
on the other, seem to provide interesting insights into the
correspondence between partial BDEs and PDEs.
In Sects.~5.1--5.3 we have only considered relatively easy cases,
in which close correspondence exists between our partial BDEs and 
ECAs. This correspondence sheds new light on the known results of
certain cellular automata. 

We summarize our results on partial BDEs in Table~2. 
In the degenerate case of all delays being equal, 
$\theta_{ij}\equiv \theta_t$ for all $i$ and $j$, partial BDEs 
are always equivalent to particular CAs in the limit of infinite size, 
and their expected behavior can be obtained from the analogy. We gave 
examples of dissipative partial BDEs corresponding to ECAs in 
different classes of Wolfram's classification \cite{Wol94}. 
In particular, a first-order approximation of the spatial derivative in
the parabolic PDE \eqref{par}, as well as a second-order approximation 
of this derivative in the hyperbolic PDE \eqref{hyp} yielded the same partial 
BDE \eqref{hyperbde2}, equivalent to ECA 150, which can behave very
differently depending on the starting data. On the other hand, 
Eq.~\eqref{hypbde} gave immediately periodic solutions for all the
starting data; hence this partial BDE is conservative. In the binary string 
formulation reviewed in Sect.~\ref{rule}, it corresponds to ECA rule 170, 
which is among the ``forbidden'' rules of Wolfram \cite{Wol94} because it 
describes asymmetric interactions. Given its simplicity, though,
Eq.~\eqref{hypbde} and hence rule 170 seem perfectly legitimate
in the present context.

More difficult situations, such as the case of nonconstant starting data in 
the initial interval, are left for future studies. 
To give an idea of the possible outcomes, consider the solution of 
Eq.~(\ref{hyper}) with initial data of the Riemann type, 
with a single jump in $u_0(z,t)$ 
at $z=0$ and $t_0=1/2<\theta_t$, {\em i.e.}, $u_0(z,t)=1$ for 
$-\infty<z \le 0$ and $0 \le t \le 1/2$, and $u_0(z,t)=0$ in the rest of the 
initial strip, $\left\{(z,t):\,0<z<\infty, \: t < \theta_t \right\} 
\bigcup \left \{ (z,t): -\infty < z \le 0, \: 1/2<t< \theta_t \right\}$.
An obvious conjecture is that the solution will still be a right- or 
left-traveling wave, depending on the sign taken in the rhs of the partial 
BDE. More generally, certain $z$-periodic IBVPs will also be well posed and, 
in the more intriguing case of Eq.~(\ref{hyperbde2}),
additive superposition will provide insight into solution behavior, which
might include certain solutions that exhibit both $z$- and $t$-periodicity.

In the parabolic partial BDE (\ref{dissor}) we conjecture instead that
the solution $u\equiv 1$ will be asymptotically stable, even for rationally 
independent $\theta_z$ and $\theta_t$ and nonconstant initial data, at least 
for large $|z |$. Conversely, $u\equiv 0$ should be asymptotically stable when
replacing the OR operator by AND as in Eq.~(\ref{dissand}); see Theorem 2.9 in 
Sect.~2.5. However, nontrivial solutions of parabolic problems could
be obtained in the presence of time-constant or time-periodic
forcing, like for PDEs.

One further step would consist of looking at the same connectives with 
different delays, {\it i.e.} different $\theta_t$ values in the variables in 
the rhs of the equations. This should in particular allow one to better 
understand the classification of partial BDEs into conservative and 
dissipative, possibly depending on the open set of delays under consideration. 
We conjecture, moreover, that ---for distinct, irrationally related delays--- 
connectives similar to the one in Eq.~(\ref{hyperbde2}) could have solutions 
of increasing complexity. If so, this would provide us with an even richer 
metaphor for evolution than either ordinary BDEs or ECAs.

\section{What Next?}
\label{whatnext}

The most promising development in the theory and applications of BDEs
seems to be the extension to an infinite, or very large, number of
variables as discussed in Sect.~4 ($n \approx 10^3$) and Sect.~5 
($n\to\infty$). Inhomogeneous partial BDEs can be easily handled by introducing
``variable coefficients,'' {\it i.e.} multiplication of the
right-hand side by site-dependent Boolean variables or functions.

Another important possibility is the randomization of various aspects of the 
BDE formulation. Wright {\it et al.} \cite{WSM90} have already considered
ensemble averaging over BDE solutions with randomized initial data, while 
Zaliapin {\it et al.} \cite{ZKG03a,ZKG03b} have considered random forcing.
It would be even more interesting to consider the random perturbation of 
delays, first in ordinary and then in partial BDEs; see also work along these 
lines in Boolean networks \cite{KlBo05}.

Next, an important but harder goal of the theory will be to develop 
{\it inverse methods}. In other words, given the behavior of a complicated 
natural system, which can be described in a first approximation by a finite 
number of discrete variables, one would like to discover a good connective 
linking these variables and yielding the observed behavior for plausible 
values of the delays. In this generality, the inverse problem for ODEs, say, is
clearly intractable. But some of the direct results obtained so far 
--- see, for instance, the asymptotic-simplification results
in Sect.~2.5 --- hold promise for BDEs, at least within certain classes and 
with certain additional conditions. Intuitively, the behavior of BDEs, 
although surprisingly rich, is more rigidly constrained than that of flows.
Certain inverse-modeling successes have also been reported for cellular 
automata; see \cite{Wol94} and references there.

From the point of view of applications, BDEs have been applied fairly 
extensively by now to climate dynamics 
\cite{GMP87,DM93,WW95,WSM90,Nic82,MMM90,Kel83} 
and are making significant inroads into solid-earth geophysics 
\cite{ZKG03a,ZKG03b}. Most interesting are recent applications
to the life sciences (Neumann and Weisbuch \cite{NW92a,NW92b},
Gagneur and Casari \cite{GC05}, \"Oktem {\it et al.} \cite{Oktem03}),
which represent in a sense a return to the motivation of the geneticist 
Ren\'e Thomas, originator of kinetic logic \cite{Thom79,Thom73,Thom78};
see also Sect.~2.6 and Appendix~A for details.

BDEs may be suited for the exploration of poorly understood phenomena in the 
socio-economic realm as well. Moreover, the robustness of fairly regular 
solutions in a wide class of BDEs, for many sets of delays and
a variety of initial states, suggests interesting applications to certain 
issues in massively parallel computations.

{\bf Acknowledgements.}
It is a pleasure to thank the organizers of the
International Workshop and Seminar
``Dynamics on Complex Networks and Applications'' (DYONET 2006),
especially J\"urgen Kurths,
for the opportunity to interact with the participants,
as well as with the members of the
Max-Planck-Institut f\"{u}r Physik Komplexer Systeme in Dresden, Germany,
and prepare this research-and-review paper. 
We are indebted to all the collaborators who helped us 
formulate, analyze, and apply BDEs:  
D. P. Dee, 
V. I. Keilis-Borok, 
A. P. Mullhaupt, 
P. Pestiaux, and 
A. Saunders.
Comments and suggestions by E. Simonnet and G. Weisbuch
 have helped improve the original manuscript. 
This work was supported by NSF Grant ATM-0327558, 
by the European Commission's Project no. 12975 (NEST) "Extreme Events: 
Causes and Consequences (E2-C2)," and by the "R\'eseau francilien
de recherche sur le d\'eveloppement soutenable (R2D2)" of the
R\'egion Ile-de-France.

\section*{Appendix~A: BDEs and Kinetic Logic}
A mathematical model closely related to BDEs was formulated
by R. Thomas \cite{Thom73,Thom78} in a slightly different form.
The difference can best be explained in terms of ``memorization 
variables'' $x_{ij}(t)=x_j(t-\theta_{ij})$.
In our formulation $x_{ij}(t)$ is a purely delayed state variable; that
is, $x_{ij}(t)$ is determined only by the state of the system
at time $t-\theta_{ij}$.
Thomas allows the memorization variable $x_{ij}(t)$ to depend
on the state of the system up to and including time $t$: when a change takes 
place in any variable $x_k$ in the time interval $(t-\theta_{ij},t)$,
then the memorization variable $x_{ij}$ may change as well.

Specifically, if for some $t'\in (t-\theta_{ij},t)$ the state of 
the system is such that
\begin{equation}
x_j(t')=f_j(x_1(t'),\dots,x_n(t'))\ne x_j(t-\theta_{ij}),
\label{kla}
\end{equation}
then the memorization variable $x_{ij}(t)$, previously equal to
$x_j(t-\theta_{ij})$, is changed to
\begin{equation}
x_{ij}(t)=x_j(t').
\label{klb}
\end{equation}
The adjustment of the variables $x_{ij}$ is in effect selectively 
erasing some of the memory of the system. 
The resulting solutions are usually simpler and hence easier to
study than the typical solution of our BDEs.
Referring to the examples shown in Figs.~2 and 3 here,
the increasing complexity of the solution reflects the fact
that the memory of the system contains more and more 
information as time goes on.

In Thomas's kinetic-logic formulation, such solutions 
of increasing complexity cannot arise.
They are due precisely to the ``conflicts'' between variable
values that were avoided on purpose in kinetic logic
\cite{GC05}. 
Some recent work on Boolean networks \cite{KlBo05,Oktem03} use a similar 
policy of eliminating behavior that appears to be too
complicated, at least for certain purposes, and thus too hard to capture 
numerically.
Eliminating the selective memory erasure of
Eqs.~\eqref{kla} and \eqref{klb} seems, on the other hand, to provide
a cleaner, richer and more versatile mathematical theory
\cite{Dee84,Mull84,GhilMull85}.

\newpage
\newpage

\begin{center}
{\bf Table 1.} Common Boolean operators
\end{center}
\medskip
\begin{tabular}{||c|c|c|c||}
\hline
\hline
Mathematical & Engineering & Name & Description\\
symbol       & symbol      &      &        \\
\hline
\hline
$\oo x$      & NOT $x$     & negator &     
not true when $x$ is true\\
$x\vee y$    & $x$ OR $y$  & logical or & 
true when either $x$ or $y$ or both are true\\
$x\wedge y$  & $x$ AND $y$ & logical and &
$x \wedge y\equiv\oo{(\oo x\vee \oo y)}$\\
$x\bigtriangledown y$ & $x$ XOR $y$ & exclusive or &
true only when $x$ and $y$ are not equal\\
$x\bigtriangleup y$ & & & true only when
$x$ and $y$ are equal\\
\hline
\hline
\end {tabular}

\newpage

\begin{center}
{\bf Table 2.} Results on partial BDEs
\end{center}
\medskip
\begin{tabular}{||c|c|c|c|c|c||}
\hline
\hline
PDE & Approxim- & PBDE
& ECA  & ECA  & Eq. $n^0$ and 
%\vspace{-.5cm}
\\ 
 $\partial_t v=$ & ation &  $u_i(t+\theta_t)=$ & rule & class & behavior \\
\hline
\hline
$\partial_z v$ &
 $\bigtriangledown$ (1st order) & 
$u_{i-1}(t)$ &
170 &
--- &
\eqref{hypbde} C \\ \hline
$\partial_{zz} v$ & 
 $\vee$ & 
$[u_{i-1}(t)\vee u_{i+1}(t)]\bigtriangledown u_i(t) $ &
54 &
I &
\eqref{dissor} D \\ \hline
$\partial_{zz} v$ & 
 $\wedge$ & 
$[u_{i-1}(t)\wedge u_{i+1}(t)]\bigtriangledown u_i(t) $ &
108 &
II &
\eqref{dissand} D
\\ \hline
$\partial_{zz} v$ & 
 $\bigtriangledown$ (1st order) &&&& 
%\vspace{-1cm}
\\ 
%\vspace{-1cm}
&& $[u_{i-1}(t)\bigtriangledown u_{i+1}(t)]
\bigtriangledown u_i(t) $ &
150 &
III &
\eqref{hyperbde2} D \\ 
$\partial_{z} v$ & 
 $\bigtriangledown$ (2nd order) &&&&\\  
\hline
\hline
\end{tabular}

\vspace{1cm}

Summary of results on the partial BDEs (PBDEs) obtained from different 
approximations for the spatial derivative in the simplest parabolic and 
hyperbolic PDEs. The temporal derivative is always approximated 
to first order by the $\bigtriangledown$ operator. 
The last column gives the equation number and the behavior of the solutions,
conservative (C) or dissipative (D).
Notice that, though all but Eq.~\eqref{hypbde}
in this table are dissipative, it is only 
Eq.~\eqref{hyperbde2} that displays chaotic behavior in the
limit of infinite lattice size.

\newpage

\end{document}